\documentclass{ws-p8-50x6-00}

\def\ltsim{\raisebox{-.5ex}{$\;\stackrel{<}{\sim}\;$}}
\def\gtsim{\raisebox{-.5ex}{$\;\stackrel{>}{\sim}\;$}}

\begin{document}

\title{Accretion and emission processes in AGN}

\author{S. Collin}

\address{Observatoire de Paris-Meudon, 92195 Meudon, France\\
E-mail: suzy.collin@obspm.fr}  


\maketitle

\abstracts{The UV-X continuum, the X-ray spectral features, 
and the variability in these bands provide powerful tools for 
studying the innermost regions of AGNs from which we gain
an insight into the accretion process.  In this chapter the discussion 
focusses on luminous AGN, i.e. Seyfert 
galaxies and quasars. The standard accretion disk 
model (a stationary geometrically thin disk) is described, and 
vertically averaged solutions for the radial 
structure are given. The emission of the standard disk is discussed using 
different approximations, and it is compared to the
observations.  This leads to the conclusion that more complex 
models are required, such as the irradiated disk and the disk-corona models. The 
advantage of this last model is that it explains the overall UV-X spectral 
distribution. In the framework of these disk models, the 
profile, intensity, 
and variability properties of the X-ray iron line can be explained by reprocessing at 
the surface of the cold disk very close to the black hole (the ``relativistic 
disk model"). An alternative possibility is discussed, where the UV-X continuum
is produced by a quasi-spherical distribution of dense  
clouds surrounded by (or embedded in) a hot medium. In such a model the iron 
line profile could be due to Comptonization instead of
relativistic effects.}

\section{A Few Introductory Words}

 In these lectures, I adopt the ``black hole paradigm",
in which AGN are 
fueled by accretion onto a massive black hole (BH)\cite{Ree84}.
Besides the BH paradigm, there is another generally accepted model, 
i.e. that accretion takes place via a disk. 
A few initial
comments should be made concerning this point.

One tries to
account for AGN luminosity by invoking accretion with 
angular momentum. If the  
flow is rotationally supported, it will form a disk. The 
presence  of a disk is inferred from several observational 
facts.  The most obvious one is the existence of a privileged 
direction: several active galaxies display cones of 
ionized gas, and collimated jets are commonly observed in radio-loud objects, 
as well as in a few radio-quiet ones. The ``Unified Scheme"\cite{AnMi85}
 implies 
the existence of a dusty torus.  Double peaked line 
profiles observed in a small fraction of AGN are also 
believed to be the signature 
of disk emission. Finally, large gaseous disks have been observed in some 
cases, for instance in NGC 4258.  
However, one should keep in mind that all these 
evidences point towards the existence of a disk {\it at large distances}, 
i.e. at $R \gtsim 10^4 R_{\rm G}$ ($R_{\rm G}=GM/c^2$ is the gravitational radius, 
where $M$ is the BH mass), while the bulk of the 
accretion luminosity is emitted by a very small region, at $\sim 10R_{\rm G}$.
 At such small distances the only evidence for an 
accretion disk is the profile of the Fe~K line. Since this 
issue is important (as it implies the presence of a ``cold disk" at only a few 
$R_{\rm G}$ in some cases), one 
should question this ``proof", and examine whether there is 
not another possible interpretation. Moreover, 
there are 
some theoretical hints for quasi-spherical accretion. For these reasons I 
discuss quite extensively the ``quasi-spherical"
model, which does not 
preclude the presence of angular momentum, although this 
is presently not a popular model. 

In order 
to show the physical scale of 
 the region discussed in this 
 chapter, Fig. \ref{fig-disque} displays a very schematic 
 representation of the central flow in an AGN.  We shall
deal with the region located inside the ``gravitationally unstable" 
part of the 
disk, closer than a few $10^3R_{\rm G}$ (the distance 
where the broad lines are formed). There is presently no 
 model proposed for the locally unstable region indicated in the figure, 
 as the gravitational torques discussed in F. 
 Combes'  lectures concern the outermost parts of the 
 nucleus, beyond 10~pc. 

\begin{figure}[t]
\begin{center}
\epsfxsize=4.5in
\caption{Schematic view of an AGN, 
showing the dimensions of different 
regions  in $R_{\rm G}$ (and in pc for $M= 10^8$ M$_{\odot}$). The present chapter 
deals with the region located inside the ``gravitationally unstable" part of the 
disk, closer than a few 10$^3R_{\rm G}$ from the BH.}
\label{fig-disque}
\end{center}
\end{figure}

I have chosen to discuss extensively
 the most simple models, and to mention  only briefly the recent more 
 sophisticated  developments. My aim is also to stress
the uncertainties underlying some widely accepted and 
 fashionable models. Finally I should mention that I focus on 
radio-quiet objects, where accretion processes dominate the spectrum.

The observational evidence is summarized 
in 
Sec.~2, and the basic physics of BHs are given in Sec.~3. 
In Sec.~4 a 
few results on the physical
 state of the medium 
emitting the bulk of the accretion power are deduced directly 
from the observations. 
In Sec.~5,  the basis of accretion disk theory is recalled, 
emphasising some robust results concerning the emission spectrum, 
which are almost 
independent of the model. In Sec.~6, the standard (i.e. thin and steady)
 $\alpha$-disks are discussed in their 
simplest (vertically averaged) form. Sec.~7 is devoted to 
the vertical structure. Sec.~8 tackles different models of 
``irradiated" disks, and their emission spectrum. Sec.~9 discusses 
the ``cloud model", which is an alternative to these disks. In Sec.~10 
the ``thick" disk models envisioned 
for very low and very high accretion rates
are briefly 
recalled.  Sec.~11 reminds us that 
all this chapter deals with regions close to the BH, and 
that the accretion process is still not understood on larger scales.

\section{Summary of Observations}

\begin{figure}[t]
\begin{center}
\epsfxsize=4in 
\caption{Typical AGN continuum of radio-loud and radio-quiet 
objects, and possible emission mechanisms (some of them not mentioned in 
the text). Figure adapted from 
Koratkar \& Blaes$^{57}$, \copyright 1999, Astr. Soc. of the Pacific, 
reproduced with permission of the editors.}
\label{fig-IR-UV-X-continuum}
\end{center}
\end{figure}


\begin{figure}[t]
\begin{center}
\epsfxsize=2.5in 
\caption{Mean line profile of the Fe~K line for a sample of Sy~1 
nuclei. The solid 
line is a double-Gaussian fit to
the profile. From Nandra et al.$^{89}$.}
\label{fig-ironKmean}
\end{center}
\end{figure}

I give here a very brief summary of some basic observational results which are 
crucial to constrain a model of the central engine. For more details the 
reader is referred to Koratkar \& Blaes' excellent review\cite{KoBl99}. 

\subsection{The Broad-Band Spectrum}

Fig.~\ref{fig-IR-UV-X-continuum} displays a
typical broad-band continuum, from the radio to the $\gamma$-ray range,
for radio-quiet and radio-loud AGN\cite{San_ea89}. 
The continuum in radio-quiet objects can be divided into 
three parts: an ``IR-bump", a ``UV-bump" (creating a gap
at about 1$\mu$m), and an X-ray power-law continuum. 
The ``1$\mu$m gap" is clearly due to the superposition of two 
different components which dominate  the IR and the UV emission,
and fade at 1$\mu$m.
Since 1$\mu$m is the shortest wavelength at which hot dust
can radiate, the IR-bump is 
attributed to hot dust radiation, at least in radio-quiet objects.


{\it The continuum produced directly by accretion processes is 
restricted to the UV-X range} (though the mid-IR continuum could be partly  
reprocessed  nuclear radiation). 
 From 1 to 10 eV the spectral index 
$\alpha$, defined as 
$F_\nu\propto \nu^{-\alpha}$, lies in the range from 
$-0.3$ (for luminous quasars) 
to +0.5 (for Seyfert galaxies and  
low  luminosity quasars). 
The  soft X-ray continuum below 1 keV seems to be simply the extension of the UV with 
a spectral index 
close to 1.5. The continuum from 1 eV 
to about 100 eV is known as the {\it Big Blue Bump} (BBB), with a peak
at 10 eV (see Fig.~5 in H. Netzer's chapter). A noticeable fact concerning the 
UV continuum is the 
almost complete {\it absence of a Lyman
discontinuity} (a weak absorption edge is sometimes observed but it is most 
probably due to the Broad Line Region or
to intervening intergalactic clouds with a redshift close to the emission 
one). I will not discuss here the ``Warm 
Absorber", which imprints features in the soft 
X-ray continuum,  as it is covered in H. Netzer's lecture notes.

 Above 1 keV the average
spectral index is about 0.7, so that in the majority of 
objects the continuum 
below 1 keV shows
 a ``soft X-ray excess"
 when compared to the extrapolation of the
continuum in the 1-10 keV range.
Recent 
observations performed with Chandra show that the strength of 
the excess has probably been overestimated in the past. Above 1 keV the spectrum can 
be decomposed into an
underlying power-law with a spectral index (in photon number) 
$\Gamma \sim 1.9$ 
and, superimposed 
on it, a broad emission line peaking at 6.4 keV, identified with
the K$\alpha$ line of weakly ionized iron, and a ``hump'' peaking at about 
30 keV (for a
review see Mushotzky et al.\cite{MDP93}). It has been 
shown\cite{ZLS99} that the strength of the hump, 
measured by a parameter $R$, correlates 
with $\Gamma$. The power-law continuum has a cut-off 
with an  e-folding energy of 50 to 300 keV.

The Fe~K line displays an asymmetrical profile 
with  an extended red wing and a smaller blue wing\cite{Nan_ea97} 
(see  Fig.~\ref{fig-ironKmean}). 
The presence of the broad red wing in the average 
spectrum has been recently questioned\cite{ZdLu00}. However in the two best 
studied 
objects, \hbox{MCG-6-30-15} and 
NGC 3516, the red wing extends down to 4 keV. The strength of the line seems also 
to correlate with $\Gamma$ \cite{ZdLu00}.

Finally, another important result is that the BBB luminosity is larger 
than the hard
X-ray luminosity, in particular in 
luminous quasars.

\subsection{Variability Properties}

Since B. Peterson focussed mainly 
on the optical-UV properties, it is necessary to recall briefly some
aspects of X-ray variability relevant to this chapter. 

A well established result is that
the characteristic time variability in the UV scales with 
luminosity: in Seyfert galaxies
it is typically of the order of a few days, while it is several months in high
luminosity quasars. The X-ray flux varies more 
rapidly 
than the UV, with typical time-scales of hours for Seyfert galaxies.

The conclusions one can draw from correlations between continuum bands 
are 
complex and sometimes controversial.
For instance, NGC 7469 shows 
variations in the UV and X-rays with similar large
amplitudes but with a time-delay of the order of 2 days, the UV leading the 
X-rays at maxima. The X-ray flux
 displays in addition short term variations not seen in the UV\cite{Nan_ea98}.
Moreover, the
X-ray spectral index seems to be correlated with the UV flux\cite{Nan_ea00}.
In NGC 3516 the delay between UV and X-bands 
is larger than 2 days\cite{Ede_ea00}.
In NGC 4051 no optical variations were observed during 
strong X-ray variations\cite{Don_ea90}.
 The UV variations drive the  optical 
variations with a time-delay $\le 0.2$~days in NGC~4151\cite{EAC96}, 
$\le 0.15$~days in NGC 3516\cite{Ede_ea00}, and $\sim 1$~day 
(but increasing with 
wavelength\cite{Wan_ea97}$^,$\cite{Col_ea98}) in NGC 7469.
Soft X-ray variations are 
generally larger than hard X-ray ones, but 
this is not always the case\cite{Nan_ea97}. In NGC 3516, they were 
strongly correlated with no measurable lag ($\le$ 0.15d)\cite{Ede_ea00}. 

The centroid energy, the intensity, and the equivalent width
of the Fe~K$\alpha$ line, are variable on short time-scales (down 
to ks). Correlations between the continuum and Fe~K
 have been intensively looked for in a few objects with ASCA, RXTE, and now 
with Chandra, revealing a 
complex behavior (for a 
review see Nandra\cite{Nan00}). ASCA 
has observed rapid (1 to 10~ks) variations of Fe~K  in  
profile and intensity in {\hbox{MCG-6-30-15}\cite{IFR96}. This 
result has been questioned\cite{Lee_ea00}$^,$\cite{CRB00}
by recent observations of \hbox{MCG-6-30-15} and NGC~5548 with RXTE, which 
show that the line flux is  constant over time-scales of 50 to 500~ks, while the profile is variable
and the underlying continuum displays large 
flux and spectral variations! The 
same result seems to be observed in NGC 3516. 
Reynolds\cite{Rey00} excludes a time-delay of 0.5 to 50~ks between the line 
and the continuum, and suggests that the line flux remains 
constant on 
time-scales between 0.5 and 500~ks in \hbox{MCG-6-30-15}. 
An analysis of a sample of Seyfert nuclei from the ASCA archive\cite{WGY00} 
finds that in most 
cases changes in the line do not appear to track changes
in the continuum. Unfortunately, the new results found with Chandra are not 
yet published at the time of writing these lectures.

 	We summarize here the most important points that any model 
should be able to explain:

\begin{itemize}
\item Spectral features:
\begin{itemize}
\item the overall shape of the continuum: the BBB
(but does it include or 
not a soft X-ray excess?), the absence of 
the Lyman discontinuity, the UV/X ratio, the X-ray hump, the 
power-law, the break at $\sim$ 100 keV, the Fe K line;
\item the correlation between $R$ and $\Gamma$;
\item the very broad red wing of the iron line (at least in a few 
objects), and its centroid energy.
\end{itemize}
\item Variability properties:
\begin{itemize}
\item the rapid variations  ($\leq$ ks) of the 2-10keV continuum;
\item the correlation between (and only between) long-term variations of UV 
and X-ray fluxes;
\item the very short (or absent) time-lag between optical and UV variations.
\end{itemize}
\end{itemize}
And perhaps a few other, but not so well established, 
variability properties:
the slow variation of $R$;
the rapid changes of the Fe K profile;
and the absence of correlation of the Fe~K flux with the underlying 
continuum.

Several other issues are not mentioned here, including the polarization 
properties, in particular, the 
abrupt increase of polarization level observed at 750 \AA $\ $ in quasars.

\section{Black Holes, Accretion Rates, and Luminosities}

It is beyond the scope of this chapter to discuss 
BH physics, for 
which excellent text books already exist\cite{Wei72}$^,$\cite{MTW73}.
For the purpose of this chapter it is only 
necessary to know that there are two types of BHs: non rotating 
or Schwarzschild BHs; and rotating or Kerr BHs.  The innermost 
stable  orbit of a Schwarzschild BH is
equal to $6R_{\rm G}$, while it depends on the 
angular momentum, $J$, for a Kerr BH. 
Due to photon capture the maximum 
value of $cJ/GM^2=0.998$ and not unity, and it corresponds to a radius of the last stable 
orbit equal to 1.24$R_{\rm G}$. Inside the innermost stable orbit, a particle 
falls almost  radially onto the BH. 
The binding energy of the last 
stable orbit determines the efficiency of energy conversion.
For Schwarzschild and Kerr BHs this is $\eta=0.057$ and
$\eta \sim 0.30$  respectively.

Another important parameter is the ``gravitational redshift". 
This is due to the
increase of proper-time with decreasing distance from the BH. Since $r=R/R_{\rm G}$
can be smaller for Kerr BHs, photons orbiting around a Kerr
BH can be more
redshifted than around a Schwarzschild BH.

The Eddington luminosity, $L_{\rm Edd}$, describes the maximum available
power due to accretion. It
corresponds to the equality between the 
gravitational force 
exerted on protons and the radiative force.
In a fully ionized gas, the
opacity is due to Thomson scattering, and this leads to the well known 
expression
\begin{equation}
L_{\rm Edd}={4\pi\ c\ GM\mu_e \over \sigma_{\rm T}} \mbox{\ \ ,}
\label{eq=eddingtonbis}
\end{equation}
where $\sigma_{\rm T}$ is the Thomson cross section 
and $\mu_e$ is the unit mass per electron.
In units appropriate for quasars, this can be written as
\begin{equation} 
L_{\rm Edd}=1.5\times 10^{46} M_8\ \ {\rm erg\ s}^{-1} \mbox{\ \ ,}
\label{eq=eddingtonter}
\end{equation}
where $M_8$ is the mass expressed in 10$^8$ M$_{\odot}$.
Note that this expression is valid in principle only for 
spherical accretion of a 
fully ionized medium in steady state.

The corresponding accretion rate is
\begin{equation}
\dot{M}_{\rm Edd}={L_{\rm Edd}\over c^2\eta} = 2.8\ M_8\left({\eta\over
0.1}\right)^{-1} \ \ {\rm M_{\odot} yr^{-1}} \mbox{\ \ .}
\label{eq=accretion}
\end{equation}
We will often use 
the ``Eddington ratio" $\dot{m}=\dot{M}/\dot{M}_{\rm Edd}=L/L_{\rm Edd}$. 
Some people define
instead $\dot{m}=c^2\dot{M}/L_{\rm Edd}$, which differs by 
$\eta$ from our expression.

B. Peterson showed in his lectures that BHs in 
Seyfert nuclei have a mass range of
10$^7$ to 10$^8$ M$_{\odot}$, with Eddington ratios of about 0.1.

\section{What Does the BBB Tell us About the Emitting Medium?}

The shape of the BBB implies thermal 
emission from an optically thick 
medium radiating 
as a black body (BB) at temperatures from a few 10$^4$K (optical emission) 
to a few 10$^5$K (EUV emission). The medium can however be optically thin at 
frequencies larger 
than the peak of the local Planck curve. 

If the medium radiates like a BB, the size of the emitting region,
$R_{\rm BBB}$, is given by
\begin{equation}\label{eq=size} 
L_{\rm BBB}= {\rm Min}(1,\tau)\ a\ 4\pi R_{\rm BBB}^2\ \sigma T_{\rm eff}^4\ 
\mbox{\ \ ,}
\end{equation}
where $L_{\rm BBB}$ is the luminosity 
of the BBB, $\tau$ is the optical thickness,  $a$ is a factor 
of the order of unity 
that depends on the geometry of the emitting region (in the 
case of a disk $a=0.5$), $\sigma$ is the Stefan constant, and 
$T_{\rm eff}$ is the effective 
temperature. Since $\tau$ is larger than unity, 
using the definition of $R_{\rm G}$
one gets 

\begin{equation}\label{eq=sizebis} 
r_{\rm BBB}={R_{\rm BBB}\over R_{\rm G}}\sim 30\  a^{1/2}\ T_5^{-2}
 M_8^{-1/2}\left({L_{\rm BBB}\over L_{\rm Edd}}\right)^{1/2}
\mbox{\ \ ,}
\end{equation}
where $T_5$ is the effective temperature in units of 10$^5$K.

One can deduce that the density of the emitting medium
 is  high. Let us assume that the UV emitting medium ``sees" the X-ray
 source, and that the surface layers are heated by this source. 
The temperature $T$ of an 
X-ray heated gas, assuming a spherical geometry,
is governed by a parameter $\Xi$ defined as

\begin{figure}[t]
\begin{center}
\epsfxsize=2.5in
\caption{$T$ vs. $\Xi$ for an optically thin medium exposed to a mixture of
 power-law continuum (of index 
0.8 in $F_{\nu}$) and a 10 eV BB, for increasing fractions of the 
BB intensity from top to bottom. From Nayakshin \& Kallman$^{96}$.}
\label{fig-Xi-T}
\end{center}
\end{figure}

\begin{equation}
  \Xi\ \sim  {1\over n_eckT}\ {L_X\over 4\pi R^2} \mbox{\ \ ,}
\label{eq-Xi-T}
\end{equation}                                                                 
where $L_{\rm X}$ is the X-ray luminosity. Note that $\Xi$ is closely related to the 
ionization parameter $U$ used by H. Netzer in his lectures.
The $\Xi$ vs. $T$ curve depends on 
the incident spectrum, as first shown by Krolik et al.\cite{KMT81}. 
 This curve is characterized by two thermally stable 
solutions: a cool phase up to
 about  $2\times 10^5$ K, where the energy 
balance is due to atomic processes;
and a hot phase close to the 
``Compton temperature" $T_{\rm Comp}$, 
that corresponds to the balance between Compton 
heating and Compton cooling.
As an illustration, Fig.~\ref{fig-Xi-T} displays the $\Xi$ vs. 
$T$ curves for an 
optically thin gas illuminated by a mixture of power-law continuum (index 
0.8 for $F_{\nu}$) and a 10 eV BB, for increasing fractions of the 
BB intensity.  One can derive that the fraction 
of soft photons decreases with
an increasing $T_{\rm Comp}$. For instance, the composite Laor et 
al\cite{Lao_ea97} AGN 
continuum has $T_{\rm Comp}=2.5\times 10^6$ K, corresponding to 
the lower curve on the figure.
 We shall come back to the $\Xi$ vs. $T$ curve
in the following sections. For the moment, we only need to know 
that whatever the fraction of BB emission, 
 a temperature of 
10$^5$K is obtained for  $\Xi   
\sim 1-10$.

Since $\Xi$ must be of the order of  a few 
units, using Eq.~(\ref{eq=size}) one gets:
\begin{equation}
n_e \sim 3\times{10}^{16}\ T_{5}^{-1}\ \ M_8^{-1}\ \left({L_{\rm X } \over  
L_{\rm Edd}}\right)
\left({R \over 10{R}_{ G}}\right)^{-2}\ {\rm cm}^{-3} \mbox{\ \ .}
\label{eq=nR} 
\end{equation}
Of course, things are more complicated if the medium is stratified and 
heated by non radiative processes, but this result remains basically correct.

The suggestion that this thick, dense, and relatively cold medium emitting 
the  BBB could be 
identified with an accretion disk was made for the first time by 
Shields\cite{Shi78} 
for 3C~273, followed by subsequent papers\cite{MaSa82}$^,$\cite{Mal83}.  
There are however other alternative explanations which will be 
discussed at the end of the chapter. We now concentrate on the physics 
of ``standard accretion disks'' and their emission properties.

\section{Thin Steady Accretion Disks: the Simplest Approximation}

First we can show that	{\it if the disk radiates like a BB at 
$\sim 10^5$K, it is geometrically
 thin}, i.e. its scale height $H$ is much smaller than its 
radius $R$. The vertical component of the gravitational acceleration at an 
altitude $z$ is $GMz/R^3$, so $H \sim R c_{\rm s}/ V_{\phi}$, 
where  $c_{\rm s}$ is the sound velocity and $V_{\phi}$ the orbital 
velocity. Although the temperature inside the 
disk can be (and 
actually is) larger than 10$^5$K, 
at 10-100$R_{\rm G}$ \ \ $c_{\rm s} \ll V_{\phi}$, which equals a fraction of the 
light velocity, so $H/R \ll 1$. 

Since the disk is geometrically thin, the problem is 
reduced to two dimensions. The gas velocity has only two 
components, $V_{\phi}$ and $V_r$, the radial component. 
If the BH mass dominates the mass of the disk and the central stellar 
cluster (which is the case for $R<10^4 R_{\rm G}$),  $V_{\phi}$ is 
Keplerian, so the orbital frequency $\Omega_{\rm K} =(GM/ 
R^3)^{1/2}$, and one deduces the basic equation
\begin{equation}
H\sim c_{\rm s} \Omega_{\rm K} \mbox{\ \ ,}
\label{eq-basic} 
\end{equation}
which is a simplified form of the hydrostatic equilibrium equation.

At this stage we can determine the 
dynamical time $t_{\rm dyn}$, i.e. the time it takes for a perturbation to 
cross the disk vertically. According to Eq.~(\ref{eq-basic}),
\begin{equation}
t_{\rm dyn}\ = \ {H\over c_{\rm s}} = \ {1\over \Omega_{\rm K}} \mbox{\ \ ,}
\label{eq-tdyn}
\end{equation}
which written in appropriate units gives
\begin{equation}
t_{\rm dyn}\sim 500 M_8r^{3/2}\ \ {\rm s.} 
\label{eq-tdynbis}
\end{equation}
This is the shortest time on which one can expect a change of structure to occur at a 
given radius.

The surface density, $\Sigma$, and $V_r$ are governed by the 
continuity equation and the equation of motion:
\begin{equation}
{\partial \Sigma\over \partial t} + {1\over R} {\partial\over 
\partial R}( R\Sigma V_r ) = 0 \mbox{\ \ ,}
\label{eq-cont} 
\end{equation}
\begin{equation}
{\partial \over \partial t}(R^2\Sigma\Omega_{\rm K}) + {1\over R} {\partial 
\over \partial R}(R^3\Sigma\Omega_{\rm K}   V_r ) = {1\over 2\pi R} {\partial \cal{G}\over \partial R} \mbox{\ \ ,}
\label{eq-NavierStokes} 
\end{equation}
where $\cal{G}$ is the torque between two adjacent rings.

To simplify the discussion, we consider a stationary disk. This  
is the case if the disk is fueled at a constant rate at its outer 
radius during the time it takes for the gas to reach the BH (the 
``viscous time"), 
and if there is no matter falling radially onto the disk at smaller 
radii.
Instabilities can however suppress the stationarity. 
Under those conditions Eq.~(\ref{eq-cont}) becomes 
\begin{equation}
\dot{M}=2\pi R\Sigma V_r \mbox{\ \ ,}
\label{eq-contbis}
\end{equation}
where the accretion rate $\dot{M}$ is constant along the radius; and 
Eq.~(\ref{eq-NavierStokes}) becomes 
\begin{equation}
{\cal{G}}=R^2\Omega_{\rm K} \dot{M}\ + \ {\rm constant} \mbox{\ \ ,}
\label{eq-NavierStokesbis}
\end{equation}
which describes the constancy of the angular momentum flux. 
From  Eq.~(\ref{eq-contbis}) it is interesting to
 note that the thin disk assumption (i.e. 
$V_r\ll V_{\phi}$) implies a large value of the surface density, 
\begin{equation}
\Sigma \sim 10^{22} {1\over \eta} {c\over V_r} {\dot{m}\over  r}\ \ {\rm 
g \ cm}^{-2}.
\label{eq-Vr-tau}
\end{equation}

Since matter orbits in circles
it requires a dissipation mechanism to lose energy and move towards the BH.
The question is how to 
ensure accretion, or in 
other words: what is the torque $\cal{G}$?
  Microscopic viscosity is far too small, 
so  one has to seek for a more 
 efficient mechanism.  Let us assume for the moment that there is a {\it 
 local}  unspecified 
viscosity. 

The kinematic viscosity $\nu$ is related to the torque through the viscous 
stress  tensor, giving 
the rate at which angular momentum is transported. For a thin disk, the 
only interesting coordinate of the stress tensor is
\begin{equation}
T_{r\phi}=\rho \nu R {\partial \Omega_{\rm K} \over \partial R} \mbox{\ \ ,}
\label{eq-alfabis}
\end{equation}
where $\rho$ is the density. The torque is thus equal to
\begin{equation}
{\cal{G}} = \int {R d\phi} \int {R T_{r\phi} dz}\ = \ 2\pi R^3 \nu \Sigma 
{\partial \Omega_{\rm K} \over \partial R} \mbox{\ \ .}
\label{eq-alfater}
\end{equation}
Note that 
since $\partial \Omega_{\rm K} / \partial R$ is negative,
 the torque is negative, and 
angular  momentum is transported outwards, as required for accretion 
to take place. 

Finally, one can compute the rate of energy dissipated in the disk per unit 
surface, $D(R)$. According to our assumption of a local viscosity, 
the potential energy is
transformed into heat, so we would get $2\pi D(R)=GM\dot{M}/R^3$. 
However, we also have to take into account the conservation of angular 
momentum and the fact that the torque cancels at $R_{\rm in}$, the radius
of the last stable 
orbit. This leads to an extra factor $(3/2) ( 1-\sqrt{ {R_{\rm in} / R}})$
(for a detailed demonstration see
Krolik\cite{Kro99}).  Finally one gets
\begin{equation}
D(R) = {3GM\dot{M}\over 4\pi R^3} f(R) \mbox{\ \ ,}
\label{eq-dissipation}
\end{equation}
where $f(R) =( 1-\sqrt{ {R_{\rm in} / R}}) $. In a fully
relativistic treatment the expression of 
$f(R)$ would be more complicated.
It is important to realize that the approximate
$R^{-3}$ dependence of the 
dissipation leads to a luminosity roughly proportional to $1/R$. 
Therefore, {\it the bulk of the gravitational energy is released close to
 the BH, 
at a few $R_{\rm G}$, but this does not mean that the bulk of the radiation
necessarily comes from this region} (cf. the ``cloud model").

\begin{figure}[t]
\begin{center}
\epsfxsize=2.1in 
\caption{Disk spectrum, assuming that the disk radiates locally like a 
BB.  Boundary corrections were not taken into account.
The index 2  and the 
exponential cut-off correspond to 
the Rayleigh-Jeans and to the 
Wien  part of the spectrum respectively.}
\label{fig-spectreBB}
\end{center}
\end{figure}

If the disk radiates like a BB, the temperature $T_{\rm BB}$  is equal 
to the effective temperature $T_{\rm eff}$. 
Since the energy is radiated by both faces of the disk, $T_{\rm eff}
=(D/2\sigma)^{1/4}$, 
which translates into 
\begin{equation}
T_{\rm eff}\sim 6\times 10^5 \eta^{-1/4}f(R)^{1/4}\dot{m}^{1/4} M_8^{-1/4} 
r^{-3/4} \ \ {\rm K,}
\label{eq-Teff}
\end{equation}
where $\dot{m}$ is defined by Eq.~(\ref{eq=accretion}). Integrating the 
Planck function over the radius and neglecting the factor $f(R)$, one obtains
 the spectrum shown in Fig.~\ref{fig-spectreBB}, where $T_{\rm in}$ and 
$T_{\rm out}$ are the 
temperatures at the inner and outer radii. 
According to the Planck law, at a frequency $\nu=kT_{\nu}/h$, the 
regions with 
$T\ <\ T_{\nu}$, i.e. $R\ >\ R(T_{\nu})$, contribute negligibly to 
the emission. On the other hand, in the inner regions where $T\ >\ T_{\nu}$, 
the spectrum at the frequency $\nu$ is given by the Rayleigh-Jeans 
expression $B_{\nu}=2\nu^2c^{-2}kT_{\rm eff}$. Thus,
$L_{\nu} \propto \int_{R_{\rm out}}^{R(T_{\nu})}R^{1/4}dR$, and finally 
 {\it the emission at a frequency $\nu$ is 
dominated  by material at the radius $R(T_{\nu})$}. 

This spectrum should be corrected for relativistic effects 
close to the BH, in particular in the case of a Kerr BH, 
where the inner radius is small. The effect is to make the spectrum 
 dependent on the inclination angle, as $L_\nu$ is relativistically 
boosted towards high frequencies for edge on disks (cf.
 Fig.~\ref{fig-hubeny-Kerr-Schw}).

The fact that the 
bulk of the emission at a given wavelength is produced at the radius 
$R_{\nu}$  has important 
consequences for the variations of the different spectral
bands. Since $B_{\nu}(T)$ peaks at
$\lambda=0.51 /T$(K)~cm, one deduces from Eq.~(\ref{eq-Teff}), that the 
emission at $\lambda_{\mu}$, expressed in microns, is 
produced mainly at a distance
\begin{equation}
r\ \sim 650\ \lambda_{\mu}^{4/3}\eta^{-1/4}f(R)^{1/4}\dot{m}^{1/4} M_8^{-1/4} 
\mbox{\ \ ,}
\label{eq-distance}
\end{equation}
or
\begin{equation}
R\ \sim\ 1.2 \lambda_{\mu}^{4/3}\eta^{-1/4}f(R)^{1/4}\dot{m}^{1/4}
M_8^{3/4}\ \ {\rm lt-d.}
\label{eq-distancebis}
\end{equation}
From this we infer that the optical to near-UV emission corresponds to regions 
located at large distances from the BH. 
Comparing this result with the small time-delays  between the UV  and optical 
light curves (less than a fraction 
of a day), we see that it is marginally compatible with a causal link propagating 
at the speed of light. It is however incompatible
with the dynamical time, according to Eqs.~(\ref{eq-tdynbis}) and 
(\ref{eq-distance}).
 But before we accept this conclusion we should question whether it is legitimate to assume that the 
disk is radiating locally like a BB, and therefore we have to 
obtain information on the optical thickness of the disk, and more 
generally on its 
physical properties. In this aim we have to specify the viscosity mechanism.

\section{A Further Step: the $\alpha$-Prescription, Vertically Averaged Models}

\subsection{Disk Structure}

It is widely accepted that the mechanism for transportation 
of angular momentum 
can be identified with ``turbulent viscosity" according 
to the  ``$\alpha$-prescription" proposed by Shakura \& Sunyaev\cite{ShSu73}
(for a detailed  
discussion of $\alpha$-disks see Frank et al.\cite{FKR92}). 
In this prescription it is assumed that the {\it sizes of 
the turbulent eddies are 
smaller than the 
thickness 
of the disk and that the turbulence is subsonic}, hence
\begin{equation}
\nu=\alpha c_{\rm s} H \mbox{\ \ ,}
\label{eq-alfa}
\end{equation}
where $\alpha \ltsim 1$ .  

The $\alpha$-prescription has proven to 
give satisfying results when applied to disks in 
cataclysmic variables, where a relatively small value ($\le 0.1$)
of  $\alpha$ is generally required.
A magnetohydrodynamic instability pointed out by Chandrasekhar\cite{Cha60}
is able  
to provide this viscosity\cite{BaHa91}. 
 Other prescriptions based on 
purely hydrodynamic (shear) 
instabilities  are also under 
debate\cite{RiZa99}, in which the corresponding value of 
$\alpha$ can be larger and vary with $\Omega_{\rm K}$. 
The viscosity can be parameterized with 
a fixed Reynolds number\cite{HuHu97}$^,$\cite{HuHu98}$^,$\cite{RiZa99}, 
following an approach 
suggested already by Lynden-Bell 
\&  Pringle\cite{LyPr74}.
Finally the viscosity 
might also be a non-local mechanism, for instance in the case 
of an organized magnetic 
field   anchored 
in the disk.

From Eqs.~(\ref{eq-contbis}) 
and (\ref{eq-NavierStokesbis}), one finds
\begin{equation}
V_r ={3\over 2} \alpha c_{\rm s}{H\over R} \mbox{\ \ ,}
\label{eq-Vr}
\end{equation}
which shows that the radial velocity is strongly subsonic, and a 
fortiori strongly sub-keplerian, as expected.

The $\alpha$-prescription is 
equivalent to the previous expression if {\it and 
only if} the disk is not self-gravitating.
Eqs. ~(\ref{eq-alfabis}) and (\ref{eq-alfa})
give
\begin{equation}
T_{r\phi}={3\over 2} \alpha \rho c_{\rm s} H \Omega_{\rm K} \mbox{\ \ .}
\label{eq-alfater}
\end{equation}
Using $c_{\rm s} =H\Omega_{\rm K}$, one gets
\begin{equation}
 T_{r\phi}={3\over 2} \alpha P \mbox{\ \ ,}
\label{eq-alfater}
\end{equation}
where $P$ is the sum of the radiative and gas pressure. 

 If the ratio of the vertical component 
 of the disk gravity to that of the central gravity, $\zeta = {4\pi G\rho / \Omega_{\rm K}^2}$  is of the order of unity, 
then\cite{Hur_ea94}:
\begin{equation}
c_{\rm s}=H\Omega_{\rm K}(1+\zeta)^{1/2}\ \ \ \  {\rm and} \ \ \ \  T_{r\phi}={3\over 
2} \alpha P(1+\zeta)^{-1/2} \mbox{\ \ .}
\label{eq-alfater}
\end{equation}
Actually there is a debate about the value of $P$ one should take in the stress 
tensor: is it the total pressure or only the gas pressure? For the moment we 
adopt the use of the total pressure.

Because it is a matter of confusion, it is necessary to stress here 
 that $T\sim T_{\rm mid-plane}$ and not $T_{\rm eff}$. If the
diffusion approximation 
holds between the two temperatures, then $T\sim \tau^{1/4} T_{\rm eff}$, 
which
implies that $T\gg T_{\rm eff}$ in 
the optically thick region.

All these expressions are 
 {\it vertically 
 averaged}.
Since the disk is geometrically thin, these averaged quantities
can be used to decouple the radial and vertical structures. Additional equations describe:

\begin{itemize}
\item The gas pressure:
\begin{equation}
P_{\rm gas} = \frac{\rho k T }{\mu m_{\rm H}} \mbox{\ \ ,}
\label{eq:Pgas}
\end{equation}
where $\mu$ is the mean mass per 
particle.
\item The radiative pressure, which in the 2-stream approximation is given by
\begin{equation}
 P_{\rm rad}={{F}_{\rm rad} \over c}\left[{1 \over 2}\tau_{\rm R}+{1 
\over \sqrt
{3}}\right] \mbox{\ \ ,}
\label{eq:Prad}
\end{equation}
where $\tau_{\rm R}$ is the Rosseland optical thickness of the disk.
\item The flux radiated by each face of the disk\cite{Hub90}
\begin{equation}
F_{\rm rad}=\frac{4}{3}\frac{\sigma T^4}{ \frac{1}{2} \tau_{\rm 
R}+\frac{1}{\sqrt{3}}+\frac{1}{3\tau_{\rm P}}} \mbox{\ \ ,}
\label{eq:Frad}
\end{equation}
where $\tau_{\rm P}$ is the Planck optical depth. Note that, in the
optically thick case,  $P_{\rm rad}$ tends to its LTE value and 
$F_{\rm rad} \approx (8/3)\sigma T^4 / \tau_{\rm R}$, which is the well-known 
diffusion approximation. In the 
optically thin case, $F_{\rm rad} \approx
4\sigma T^4 \tau_{\rm P} $.
\item The energy balance between the local viscous dissipation and the
radiative cooling,
\begin{equation}
F_{\rm rad} = D(R)/2 \mbox{\ \ .}
\label{eq:eb}
\end{equation}
\end{itemize}

These equations, when solved self-consistently with the appropriate 
functions for $\kappa_{\rm R}(\rho,T)$, $\kappa_{\rm P}(\rho,T)$ and 
$\mu_{\rm P}(\rho,T)$, yield the radial disk structure in the non 
self-gravitating region of the disk for given values of $M$, $\dot{M}$, 
and $\alpha$.
One finds that the disk is divided into several regions, according to the 
dominating processes:  radiation or gas pressure; 
Thomson, Kramers, or more complicated opacity; and
self-gravity or no self-gravity.

To show the importance of the 
opacity, 
Fig.~\ref{fig-JM-SS} displays the radial dependence of several physical 
quantities as function of the radius for an 
$\alpha$-disk in the case of a typical AGN. In these computations self-gravity has 
been neglected in order to focus on the other parameters. 
Different approximations have 
been compared to a 
 2D computation with realistic opacities. We see 
that in the inner region ($R \leq 10^{16}$ cm, i.e. $r\leq 10^3R_{\rm G}$), the 
opacity is better 
approximated by Thomson scattering, while in the outer region it is 
better approximated by Kramers' opacity. Note that the radiation 
pressure dominates 
in the Thomson regime ($\beta=P_{\rm gas}/P_{\rm rad}$). 

\begin{figure}
\epsfxsize=4.5in 
\caption{From left to right and top to bottom, 
density, surface density, Rosseland mean 
optical 
thickness, scale height, equatorial temperature and radiation 
to gas pressure ratio as functions of the radius for a vertically averaged 
$\alpha$-disk. The thick solid line corresponds to a 2D model with realistic 
opacities. Courtesy of J.-M. Hur\'e.}
\label{fig-JM-SS}
\end{figure}

\begin{figure}
\begin{center}
\epsfxsize=2.7in 
\caption{Transition from radiation to gas pressure regime for a vertically 
averaged $\alpha$-disk, for $M=10^8$M$_{\odot}$. 
The figure displays $\dot{m}$ versus $r$ for different values of 
$\alpha$. Courtesy of J.-M. Hur\'e.}
\label{fig-JM-P}
\end{center}
\end{figure}

Fig.~\ref{fig-JM-P} shows that the transition between the radiation 
and gas pressure dominated regimes occurs at a larger radius for 
a larger accretion rate in Eddington units. 
This radius also increases with a decreasing viscosity 
parameter $\alpha$, 
which  corresponds to an increasing density. For 
a typical AGN ($\dot{m}=0.1$ 
and $M=10^8$M$_{\odot}$) and
$\alpha=0.1$, $R \sim 2000R_{\rm G}$. Note that it occurs in the same region
as the transition to the self-gravity regime (this is due to the rapid increase 
of density when the disk gravity begins to dominate). 

 Finally the disk becomes self-gravitating at a radius $R_{\rm sg}$ corresponding to 
 $\zeta\sim 1$. 
As shown in Fig.  \ref{fig-JM-SG}, the transition occurs 
at $R\sim 100 -1000 R_{\rm G}$. It is located further away for a larger value of 
$\alpha$, owing to the corresponding smaller value of the density.

\begin{figure}
\begin{center}
\epsfxsize=3.5in 
\caption{Transition to the self-gravity regime for a vertically 
averaged $\alpha$-disk. 
The figure displays $R_{\rm sg}/R_{\rm G}$ versus $\dot{m}$ for different values of 
$M$. All models except one are computed  for $\alpha=0.1$. 
The curves are labelled with $\log M$ in units of 
M$_{\odot}$.}
\label{fig-JM-SG}
\end{center}
\end{figure}

In summary 
{\it the non self-gravitating part of the disk  is generally reduced to the 
Thomson opacity--radiation pressure dominated regime.} Neglecting 
the  boundary 
function $f(R)$ the 
different 
parameters vary as:
\begin{itemize}
\item $\rho \propto R^{3/2}$ \ \ ,
\item $H$ = constant  \ \ ,
\item $\Sigma$ and $\tau \propto R^{3/2}$  \ \ ,
\item $T \propto R^{-3/8}$  \ \ ,
\item $\zeta \propto R^{9/2}$  \ \ ,
\item $P_{\rm gas}/P_{\rm rad} \propto R^{21/8}$ \ \ .
\end{itemize}
These relations explain why the transition from the gas pressure to the 
radiation 
pressure regime, and the transition to self-gravity, are so rapid (see the 
strong  variation 
of $\zeta$ and $P_{\rm gas}/P_{\rm rad}$ as functions of the radius on
Fig.  \ref{fig-JM-P}). It is also interesting to note 
that the gas temperature decreases less rapidly than 
$T_{\rm eff}$ (which is proportional to $R^{-3/4}$).
The scale height to radius ratio can be 
expressed in this region as
\begin{equation}
{H\over R} = 6.7 {\dot{m} \over r} \mbox{\ \ ,}
\label{eq-H/R}
\end{equation}
which shows that the ``geometrically thin" assumption does not hold for 
large $\dot{m}$; and
the Thomson thickness is given by
\begin{equation}
\tau_{\rm T} \ = \ 55 \alpha^{-1}\left({R\over 
R_{\rm G}}\right)^{3/2} \mbox{\ \ .}
\label{eq-tauTh}
\end{equation}

 A few important time-scales relevant to the external regions which emit 
at optical wavelengths, where self-gravity is not negligible, are defined
below.
The ``viscous 
 time" $t_{\rm visc}$, basically the time it takes for matter to move to the 
 center, can be expressed as (using in particular Eq.~(\ref{eq-alfater})):
\begin{equation}
t_{\rm visc} \sim \ {R\over V_r}\ \ \sim \  {1\over\alpha} \left({H\over 
R}\right)^{-2}{1\over \Omega_{\rm K} \sqrt{1+\zeta}} \mbox{\ \ .}
\label{eq-tvisc}
\end{equation}
The ``thermal time" $t_{\rm th}$, i.e. the time it takes 
to radiate the thermal 
energy, can be expressed as
\begin{equation}
t_{\rm th}\ \sim \ {PH\over D(R)}\ \ \sim \ 
{\sqrt{1+\zeta} \over \alpha 
\Omega_{\rm K}} \mbox{\ \ .} 
\label{eq-ttherm}
\end{equation}
The expression for the dynamical time $t_{\rm dyn}$ has already been given in 
Eq.~(\ref{eq-tdyn}). Correcting for self-gravity, it 
becomes
\begin{equation}
t_{\rm dyn}\ \sim {1\over \Omega_{\rm K} \sqrt{1+\zeta}} \mbox{\ \ .}
\label{eq-tdynter}
\end{equation}
It is easy to see from these expressions that
\begin{equation}
t_{\rm visc}\gg t_{\rm th}\ge t_{\rm dyn} \mbox{\ \ .}
\label{eq-ttt}
\end{equation}
For instance in the inner Thomson opacity - radiation pressure dominated 
region $t_{\rm visc}$ is given by
\begin{equation}
t_{\rm visc}= 10\  \alpha^{-1}  
M_8 \dot{m}^{-2} r^{5/2} \ \ {\rm s.}
\label{eq-tviscbis}
\end{equation}

\vspace*{0.3cm}
\noindent
{\it 6.1.1 \ \ Stability of the Disk}
\medskip

\noindent
 There are 3 main
 instabilities acting in the disk:
\begin{enumerate}
\item the thermal instability, which occurs when
\begin{equation}
\left({\partial \ln D\over \partial \ln T}\right)_{R}-
\left({\partial \ln F_{\rm rad}\over \partial \ln T}\right)_{R} > 0
\mbox{\ \ ;}
\label{eq-therm-inst}
\end{equation}
\item the viscous instability, which occurs when
\begin{equation}
\left({\partial \ln \dot{M}\over \partial \ln \Sigma}\right)_{R} < 0
\mbox{\ \ ;}
\label{eq-visc-inst}
\end{equation}
\item the gravitational instability, which occurs 
when\cite{Too64}$^,$\cite{GoLy65}
\begin{equation}
Q = {\Omega_{\rm K}^{2}\over \pi G \rho} < 1 \mbox{\ \ .}
\label{eq-Toomre parameter}
\end{equation}
\end{enumerate}

In the self-gravitating regime, where\cite{Hur98}  $\rho \propto R^3$ and 
$\Sigma \propto R^2$,  the 
 disk shrinks, rapidly  becomes strongly gravitationally unstable, and 
 unstable fragments with a typical volume $H^3$ begin 
 to collapse. 
Consequently, for   $R > R_{\rm sg}$ 
the $\alpha$ viscosity does 
not hold. We 
 stress
 that this gravitational instability is {\it local}, and does not
 require the disk mass
 to be of the order of the BH mass. The instability
corresponds simply to the tidal 
limit $4\pi\Sigma \ge MH/R^3$, or 
equivalently to $M_{\rm disk}\sim MH/R$, hence $M_{\rm disk}\ll M$. 

\begin{figure}
\begin{center}
\epsfxsize=20pc 
\caption{Viscously (left) and thermally (right) unstable regions for a 
vertically averaged 
$\alpha$-disk, with $\alpha=1$, $M=3\times10^7$M$_{\odot}$, and $\dot{m}$=0.03. The figure displays $T$ (in K) and $H$ (in cm) versus $r$. The solid line 
represent the unstable regions. From Hur\'e$^{45}$.}
\label{fig-JM-inst}
\end{center}
\end{figure}

It can be shown that the thermal and the viscous instabilities are strongly 
linked, so they generally occur  in the same region of the disk (see Fig.~
\ref{fig-JM-inst}). We see that the disk is  
both viscously and thermally unstable in a large fraction of the
inner region. This combined instability should operate on 
a thermal time-scale, which is quite small in the inner regions 
(cf. Eqs.~(\ref{eq-tdynbis}) and (\ref{eq-ttherm})). It should therefore induce
 rapid variations of EUV emission.
Moreover as $\alpha$-disks are 
 also  gravitationally unstable 
 in the outer region, one 
deduces that {\it in AGN, $\alpha$-disks are unstable almost everywhere!}

\vspace*{0.3cm}
\noindent
{\it 6.1.2 \ \ ``$\beta$-Disks"}
\medskip

\noindent
In the case where the stress tensor is
assumed to be proportional  to the gas pressure and not to the 
total pressure\cite{ClSh89}$^,$\cite{MiSh90}, the disks are known as
``$\beta$-disks". Strictly speaking they should be called  
``$\alpha-\beta$-disks"! They
are denser than $\alpha$-disks, and dominated by gas pressure even 
in the inner regions. Consequently they are not subject to the viscous/thermal 
instability, but to another instability due to ionization 
 of hydrogen, which occurs in regions where 
the   mid-plane temperature is of the 
order of 10$^4$K. This instability induces a ``limit-cycle" behavior, which is 
thought to 
explain the periodic variations of dwarf novae. In AGN it should operate on a much 
larger time-scale (10$^5$ yr), and could correspond to variations of 
4 orders of magnitude in luminosity\cite{SCK96}. 

\subsection{Where a Fundamental Ingredient, Compton
 Scattering,
  Appears}

The solution of the vertically averaged disk 
 showed that Thomson opacity dominates over free-free 
opacity in the inner regions.
Diffusions are therefore important, and modify 
the equilibrium spectrum. Radiation is thermalized for a relatively 
small  absorption optical thickness, when 
the effective optical thickness 
$\tau_{\rm eff} = \sqrt{\tau_{\rm abs}(\tau_{\rm abs}+\tau_{\rm T})}\sim                      
\sqrt{\tau_{\rm abs}\tau_{\rm T}} >1$, where $\tau_{\rm abs}$ and $\tau_{\rm T}$ are 
the absorption and diffusion optical thicknesses respectively. 
This occurs in 
the inner regions. In this case the spectral distribution is given 
by\cite{RyLi79}
\begin{equation}
I_{\nu} = 2 B_{\nu} \sqrt{{\kappa_{\rm abs}\over\kappa_{\rm T}}} \ll B_{\nu}
\mbox{\ \ .}
\label{eq-modifiedBB}
\end{equation}
For instance, if absorption is dominated by free-free processes, 
\begin{equation}
\kappa_{\rm abs} = 3.69\times 10^8 N_eN_iZ^2g_{\rm ff}{1\over \sqrt{T}}{1\over
 \nu^3}\left(1-{\rm exp}\left(-{h\nu\over kT}\right)\right) ,
\label{eq-kappa-ff}
\end{equation}
and, using Eqs.~(\ref{eq-modifiedBB}) and (\ref{eq-kappa-ff}),  one deduces
that 
the intensity is given by
\begin{equation}
I_{\nu} \propto \nu T^{1/4} 
\label{eq-RJ}
\end{equation}
in the Rayleigh-Jeans regime, where ${h\nu / kT} < 1$; or
\begin{equation}
I_{\nu} \propto \nu^{3/2}{\rm exp}\left(-{h\nu\over kT}\right)  T^{-1/4}
\label{eq-Wien}
\end{equation}
in the Wien regime, where ${h\nu / kT} > 1$.
The spectrum emitted by the innermost regions 
 will therefore  differ from a local 
BB. 

However it is not sufficient to take into account Thomson scattering, 
 as one can easily show that  Compton scattering is also important. Since 
 the gas temperature is high and we are interested in soft X-ray photons, 
 the collisions between electrons and photons are not elastic, and they 
 lose or gain energy with each scattering. The ``Compton parameter" $y$, which is the total relative change of energy 
when a photon travels through the medium, is equal to the mean number 
of  diffusions times  the relative change of energy at 
each  scattering.
The mean number of  diffusions, $N_{\rm s} = {\rm Max} (\tau_{\rm T},\tau_{\rm T}^2)
=\tau_{\rm T}^2$. In the case of a thermal non relativistic distribution of 
electrons, the relative change of energy at 
each  scattering  is given by\cite{RyLi79}
\begin{equation}
{\Delta h\nu\over h\nu} = -{h\nu\over m_oc^2} + {4kT\over m_oc^2}
\mbox{\ \ ,}
\label{eq-compton}
\end{equation}
and thus one gets
\begin{equation}
y={4kT-h\nu\over m_oc^2}\tau_{\rm T}^2 \mbox{\ \ .}
\label{eq-comptonbis}
\end{equation}

In the inner regions of the disk, $\tau_{\rm T}\sim 10^{4}$ and $T\sim 
 10^6$K, so $y$ can easily be larger than unity. 
If free-free dominates, then
$\tau(\nu)$ and hence $y$ increases with decreasing frequency, and there is a
frequency $\nu_o$ below which $y>1$. If $h\nu_o/kT \ll 1$, inverse Compton processes will 
 drive soft photons to higher frequencies, and there will be an accumulation 
 of photons at $\nu \sim 4kT/h$, at the expense of the low 
 frequency population\cite{CzEl87}$^,$\cite{WaPe88}.
Another consequence of Compton diffusion is 
 that the profile of a line 
 will be asymmetrically broadened, as it is dominated 
 by Inverse Compton diffusions for  $\nu_{\rm line} < 4kT/h$, and 
 by direct Compton diffusions for  $\nu_{\rm line} > 4kT/h$ (i.e. 
 down-scatterings leading to a broad red wing). The computation of the spectrum 
in this case is performed
with transfer codes using 
 the ``Kompaneets diffusion equation", 
 or with Monte Carlo codes.
The effect of Compton scatterings on the spectral
distribution of the disk is illustrated in Fig.  \ref{fig-fabian}, and 
at the end of the chapter we will discuss the effect on the profile of the 
iron line.

Another problem which has to be taken into account is the existence of an 
atmosphere above the disk. As a first approach\cite{KoSu84}$^,$\cite{SuMa87},
the disk was replaced by
low gravity stellar atmospheres of different $T_{\rm eff}$. However,  since 
gravity is constant in a stellar
atmosphere, while it increases with height in disks, this method
is not valid, and  led to the erroneous result 
that the spectrum should always show a strong Lyman discontinuity
in absorption.

\section{The $\alpha$-Prescription,  
Computation of the Vertical Structure and Vertical Transfer}
 
 The computation of an accurate emission spectrum requires a model of
 vertical structure with the 
 self-consistent treatment of radiative transfer. 
 Since the disk 
 is still assumed to be geometrically thin, it is again possible to 
 decouple the vertical and the horizontal structure equations.
 One 
 then  solves for each annulus  a set of 
 differential equations in $z$, instead of a set of algebraic equations. 
A simplification consists, at least in a first step, of decoupling 
the  computation of the 
emitted  spectrum from that of the 
 vertical structure, as the disk is optically thick and the spectrum forms 
 in the upper layers. The radiative transfer is then solved using the 
diffusion 
 approximation.

\subsection{Vertical Structure}

\begin{figure}
\begin{center}
\epsfxsize=4.5in
\epsfxsize=4.5in
\caption{Temperature and density structure of an $\alpha$-disk for a typical
  AGN ($M=10^8$M$_{\odot}$, $\dot{M}=0.1$M$_{\odot}$
yr$^{-1}$ or $ \dot{m}\sim 0.03$,  $\alpha=0.1$). Courtesy of J-M. Hur\'e.}
\label{fig-JM-2D}
\end{center}
\end{figure}

The equations for the vertical structure are:
\begin{itemize}
\item the hydrostatic equilibrium,
\begin{equation}
{1\over \rho}{dP_{\rm tot}\over dz} = -\Omega_{\rm K}^2z \mbox{\ \ ;}
\label{eq-hyd-eq}
\end{equation}
\item the vertical variation of the energy produced by viscous dissipation,
\begin{equation}
{dF\over dz} = {9\over 4} \rho \nu(z) \Omega_{\rm K}^2 \mbox{\ \ ;}
\label{eq-heat-flux}
\end{equation}
\item the variation of the mean optical thickness and of the surface 
density,
\begin{equation}
{d\tau\over dz}=-\kappa \rho\ \ {\rm and}\ \ {d\Sigma\over dz}=\rho 
\mbox{\ \ ,}
\label{eq-dtau-dz}
\end{equation}
where $\kappa$ is a frequency averaged opacity;
\item the variation of the radiation pressure in the diffusion 
approximation, assuming that the energy is transported only via radiation,
\begin{equation}
-{c_{\rm s}\over \mu \kappa}{dP_{\rm rad}\over dz} = F(z) \mbox{\ \ .}
\label{eq-dtau-dz}
\end{equation}
\end{itemize}

A question is raised here:
how is viscous heating deposited vertically? The  original
Shakura-Sunyaev prescription was introduced to describe vertically
averaged quantities, and it is not clear how to extend it to a $z$-dependant 
model.
According to the $\alpha-P$ 
prescription described above, it seems natural to assume that
 $\nu$ 
is constant vertically, which implies that 
the heat deposition ${dF/dz}$ is proportional to $\rho$. If the 
opacity is dominated by Thomson scattering and the pressure is dominated 
by $P_{\rm rad}$, as is the case in the inner 
region of the disk, then Eqs.~(\ref{eq-hyd-eq}), (\ref{eq-heat-flux}) and 
(\ref{eq-dtau-dz}) lead to the result that $\rho$ does not depend on $z$.
A different prescription is sometimes used for disks in dwarf 
novae\cite{MeMe94}.

Other ingredients can also be added to the model, such as convective 
transport. As an illustration, 
Fig. \ref{fig-JM-2D} displays the 2D structure of an 
$\alpha$-disk, for $\alpha=0.1$, taking into account convective and turbulent 
transport and 
self-gravity\cite{Hur00}. The parameters are those of a typical AGN
 ($M=10^8$M$_{\odot}$, $\dot{M}=0.1$M$_{\odot}$/yr, or $\dot{m}\sim 0.03$). 
Comparing the results of 2D and 1D computations, one finds that the
 results  are similar for the radial 
structure within factors of a few. The temperature in the 
equatorial plane is accurate to within 30$\%$ and the most important 
difference (by a factor $\leq 4$) is 
in the disk thickness\cite{HuGa00}.  
The 
density and the surface density, the critical radius for self-gravity, 
the ratio of gas to radiation 
pressure ratio are correctly computed in the 1D 
model (cf. Fig.  \ref{fig-JM-SS}). The 2D structure 
avoids however some multiple solutions met 
in  the 1D model (cf. Fig. \ref{fig-JM-inst}).

\subsection{Disk Spectrum}

Vertically averaged $\alpha$-disks, which have proven to be extremely 
useful in determining the radial structure of the disk, are not 
sufficient to determine the emission spectrum. 
This requires 
solving  the vertical radiative transfer in a consistent 
way with the vertical structure of the disk. 

Several authors have built consistent
 models of the vertical
structure of AGN accretion disks using various
approximations for the radiative transfer.
The first computation was performed by
Laor \& Netzer\cite{LaNe89}, who solved the radiative transfer 
in the Eddington 
approximation, with LTE,  
bound-free and free-free opacities, 
and with different assumptions for the stress tensor as well as for vertical 
heat deposition laws. They showed
in particular that the Lyman discontinuity can appear either in absorption
or in emission according to these laws.   Ross et al.\cite{RFM92} relaxed the LTE assumption and took into account  Comptonization
using the Kompaneets equation. St\"orzer\cite{St93}  solved the
radiative transfer equation in 3D, also taking into 
account the influence of self-gravity, but keeping the LTE approximation.   
Shimura \& Takahara\cite{ShTa93}$^,$\cite{ShTa95} 
 also solved the vertical 
structure together with the transfer treated with the Kompaneets equation, 
but for a stress tensor 
proportional to the gas pressure and not to the total pressure. 
St\"orzer et al.\cite{SHA94} and Shields \& Coleman\cite{ShCo94},
addressed the question of how departures from  LTE influence the Lyman
discontinuity. D\"orrer et al.\cite{Doe_ea96} solved
the vertical structure self-consistently, with LTE and no bound-free opacities,
and a pure-H atmosphere, but taking into account
 Compton scattering. Sincell \& Krolik\cite{SiKr98} solved 
 the coupled vertical structure and radiation transfer with a two 
 stream Eddington approximation, restricting the elements to 
He and H in LTE. 

 Fig.  \ref{fig-fabian} illustrates the influence of 
Comptonization\cite{RFM92}. It
shows the spectra of two 
(non-rotating) BHs, with different masses and accretion rates.
This figure illustrates two interesting results:
\begin{enumerate}
\item The soft X-ray excess cannot be reproduced. Although a 
non-negligible fraction of 
the disk luminosity is radiated in the soft X-ray range as a consequence of 
Comptonization, it is obvious that the spectral energy distribution
of AGN cannot be 
accommodated if one only assumes an extra 
power-law component exists in the soft X-ray range.  
Maximally rotating BHs
produce 
continua which are harder those of non-rotating BHs, 
but they do not solve the 
problem either.
\item The local BB gives the correct result at  optical to near-UV 
wavelengths (this would not be true for the UV band and for higher 
 masses\cite{Hub_ea00}), and confirms
{\it our previous conclusion 
that the time-delay observed between the optical and UV bands is too small compared with the 
smallest time-scale in the disk}.
\end{enumerate}

\begin{figure}
\begin{center}
\epsfxsize=5in 
\caption{Spectrum of an $\alpha$-disk around a Schwarzschild BH. Solid lines represent the
 contribution from 
different annuli and the total spectrum (with Componization). Dotted lines
represent the local BB 
approximation used for the outermost annuli.
Dashed lines represent the total spectrum with the local BB computation. From 
Ross et al.$^{108}$}
\label{fig-fabian}
\end{center}
\end{figure}

Hubeny and collaborators\cite{HuHu97}$^,$\cite{HuHu98}$^,$\cite{Hub_ea00}
solved self-consistently 
the vertical structure with an exact radiative transfer with no diffusion 
 and escape probability approximation, in the non LTE case, taking
 into account the general relativistic
transfer function. However, they did not take into account Comptonization,
 or heavy elements,
so their models probably do not correctly describe  the EUV.
They computed the most extensive set of models to date:
a grid of non-LTE disks around 
maximally-rotating Kerr and 
Schwarzschild  BHs, for a wide range
 of BH masses
and  accretion rates. Some of their results are shown in 
Figs.  \ref{fig-hubeny-Kerr-Schw}, 
 \ref{fig-hubeny-opt-UV-slope}, and 
\ref{fig-hubeny-lyman}.

Fig.~\ref{fig-hubeny-Kerr-Schw} compares the energy distributions of
disks of  
Schwarzschild  and Kerr BHs of
10$^9$ M$_{\odot}$
viewed at 37$^o$, for different  disk luminosities. 
Schwarzschild  and Kerr  
BHs produce quite similar optical spectra, except that 
in Kerr BHs the spectrum is shifted towards 
higher frequencies. 

\begin{figure}
\begin{center}
\epsfxsize=3in 
\caption{
Spectral energy distribution of an $\alpha$-disk around a
maximally rotating BH (solid lines) and a Schwarzschild BH
(dashed lines), with $M=10^9$ M$_{\odot}$
and various values of the accretion rates. The curves correspond to 
$L/L_{\rm Edd}$= 1, 1/2, and 1/4, etc., for the Kerr BH; and 
the corresponding values for the Schwarzschild BH are 5.613 times larger. From Hubeny et al.$^{43}$}
\label{fig-hubeny-Kerr-Schw}
\end{center}
\end{figure}

Fig.~\ref{fig-hubeny-opt-UV-slope} displays 1450 to 5050~\AA\ 
spectral slopes, $\beta$   (defined here as 
 $F_{\nu}\propto \nu^{\beta}$), as a function of the Eddington 
ratio for Kerr BHs of different masses. 
 The local BB models depart from real transfer models except for low 
 accretion rates.  This is due to the relatively high masses considered. 
Both sets of  models can reproduce the mean 
observed colors of the LBQS sample\cite{Fra_ea91}.
However, one should note that {\it the models cannot account 
 for the positive slopes 
 observed in the majority of Seyfert galaxies}.

Finally, Fig. \ref{fig-hubeny-lyman} shows the strength of the Lyman edge 
in a Kerr BH with different disk inclinations. The 
Lyman discontinuity is 
always more intense than observed, except for very low accretion rates or 
very large disk inclinations, not plausible in the framework
 of Unification. It is however possible that the addition of 
metal opacity (not taken into account
 in the computations)
changes this result.

\begin{figure}
\begin{center}
\epsfxsize=2.5in 
\caption{
Optical/ultraviolet spectral slope between 1450 and 5050\AA$\ $
for an $\alpha$-disk around a Kerr BH with 37$^o$ viewing angle, 
and $\alpha=0.01$. 
Solid curves: computed models; dashed curves: local
BB. From top to bottom, the curves are computed for 9
values of $M$: 1/8, 1/4, 1/2, 1, 2, 4, 8, 16, and 32 $\times 10^9$M$_{\odot}$.
From Hubeny et al.$^{43}$}
\label{fig-hubeny-opt-UV-slope}
\end{center}
\end{figure}
\begin{figure}
\begin{center}
\epsfxsize=2.5in 
\caption{Lyman edge strength as a function
of Eddington ratio for
an $\alpha$-disk around a Kerr BH with $M=10^9$ M$_{\odot}$. The curves are labelled with
 the value of $\cos i$. From Hubeny et al.$^{43}$}
\label{fig-hubeny-lyman}
\end{center}
\end{figure}

This discussion illustrates the fact 
 that the computation of disk spectra is difficult and involves 
 poorly known parameters.  Besides, several
 other parameters not mentioned here complicate the problem, such 
as the value of $\alpha$, the inner and outer radii of the disk,
the different assumptions concerning heat deposition, etc. Therefore, 
in some cases 
it  may not be a bad
 choice to use the simplest and least
 model-dependent solutions. 
Despite these uncertainties, the 
following
results can be considered as secure: 
\begin{itemize}
\item the spectral energy distribution does not fit the observations in
the soft X-ray range: another emission mechanism is required;
\item the disk spectrum in the optical to near-UV range is too flat 
compared to that of many Seyfert nuclei;
\item the computed Lyman discontinuity is too strong or has to be fine-tuned;
\item the small or absent time-lag between the optical and UV bands is not compatible with the 
viscous time-scale, or even with the dynamical time;
\item the disk is not emitting hard X-rays at all.
\end{itemize}

Additionally, we paid no attention to polarization 
properties. They are also hard to explain in the context of the 
standard disk model. 

\section{Other Disk Models}

\subsection{The ``Lamppost" Model}

\begin{figure}
\begin{center}
\epsfxsize=7in 
\hspace*{-2.0cm}
\vspace{-2.5cm}
\caption{
Sketches of several models of irradiated disks (the lamppost, the 
disk-corona, the patchy corona), and of the cloud model.}
\label{fig-modeles}
\end{center}
\end{figure}

The discussion on the standard disk model has shown that the regions emitting
at UV and optical 
wavelengths are located at large distances from each other, so 
the small time-lags between the (well correlated) optical and UV light curves 
require a causal link propagating at the speed of light. This led to 
questioning the validity of the thin steady  disk 
model\cite{All_ea85}$^,$\cite{CoCl91}, and
to propose that the disk 
was actually 
irradiated by the X-ray continuum, and emitting as a result of both 
gravitational and 
external radiative heating\cite{Col91}$^,$\cite{MMS92}$^,$\cite{RCM93}. 
The discovery of the iron line and 
the X-ray hump\cite{Pou_ea90} implied reprocessing by a cold 
medium, in agreement with this model.

There then followed the era of ``irradiated disks". These models included an
X-ray  point source located at a given height $A$ above the center of the disk,
 or a spherical source of radius $A$ around the 
BH. The former is now referred to as the ``lamppost model" (cf. 
Fig. \ref{fig-modeles}).

The flux falling on each face of the disk 
 $F_{\rm X}$ can be expressed as 
\begin{equation}
 F_{\rm X} = af_{\rm X}{L_{\rm bol} \over 4\pi(R^2+ A^2)}\cos 
\theta \sim  f_{\rm X} {L_{\rm  bol}A \over 4\pi R^3} \mbox{\ \ ,}
\label{eq-Fincbis}
\end{equation}
where 
$f_{\rm X}$ is the fraction of the bolometric luminosity emitted by this 
source, $\theta$ is the angle of the light rays to the normal,
 and $a=1/2$ for a point source 
and $a=2/3\pi$ for  a spherical source.
 The second approximation is valid for $R \gg A$, i.e. 
for the optical to near-UV emitting regions, located much further 
away than the X-ray source.
X-ray variability studies suggest that $A \sim 10 R_{\rm G}$. 
{\it Therefore, the external flux varies with the 
radius like the viscous flux.}
This external irradiation has several effects on the disk:
\begin{itemize}
\item A major fraction of X-rays is absorbed by 
photoelectric processes in the upper layers of the atmosphere, and this 
creates 
a hot  and ionized ``skin" which emits the iron line.
\item The remaining fraction is Compton scattered in this 
ionized   atmosphere and gives rise to the 
X-ray  ``hump". 
\item The absorbed fraction is reprocessed into UV, of which
about 50$\%$ is re-emitted outwards and 50$\%$ towards the interior. 
This latter fraction can   modify the internal structure of the disk. The 
effect  is however
negligible, unless $f_{\rm X}\approx 1$.

Indeed, the
condition for radiative heating to dominate the viscous
heating when the disk is optically thick 
is\cite{Hub90}$^,$\cite{Hur_ea94}
\begin{equation}
{F}_{\rm visc}\ (1+{3 \over 8}\tau_{\rm R})\ \leq {{F}_{\rm X}\over 2} \mbox{\ \ .}
\label{eq-irradiation}
\end{equation}
Since this condition is not realized, the radial structure is not modified 
(unlike in the disk-corona model).
\item The fraction of UV radiation re-emitted outwards is added to the 
emission due to the viscous flux.
\end{itemize}

The emitted spectrum has been computed using different 
schematizations.  Except for the most recent 
ones\cite{NKK00},
the models consist of a disk of
constant density in the 
vertical direction, irradiated  on the top by a power-law 
continuum
and on the bottom by a BB (or by a Wien spectrum) 
to approximate the effect of viscous heating. The transfer is treated 
 with various 
approximations\cite{FaRo93}$^,$\cite{MFR93}$^,$\cite{CoDu97}$^,$\cite{SiKr97} 
or with a Monte Carlo 
method coupled with a 
 photoionization code\cite{Zyc_ea94}.  An example is given in 
 Figs. \ref{fig-deg-ion-Fe} and \ref{fig-spec-BB-PLx300}. Fig. 
 \ref{fig-deg-ion-Fe} shows that iron is strongly ionized in the upper 
 layers. The spectrum displayed 
 in Fig. \ref{fig-spec-BB-PLx300} illustrates the effects mentioned above.
 It is computed with the code TITAN\cite{DAC00}, which solves 
 the complete non-LTE
 transfer inside lines and continuum, with photoionization and 
 thermal equilibrium, and is coupled to a Monte Carlo code to handle 
 Compton diffusion. The 
 strong UV-bump corresponds to the viscous release, to which a 
 small ($\sim 10\%$) contribution of reflected radiation is added. It is 
 interesting to note that the Lyman discontinuity, present here in 
 absorption, is quite weak, though not completely absent, as in the observed 
 spectra.  In the soft X-ray 
 range the spectrum has a slope of the order of 1.5 in flux, in 
 agreement with the observations, and displays many lines and 
 edges from a highly ionized medium. Such lines are observed in the 
 Chandra spectra of NGC 3783\cite{Kas_ea00}, and NGC 5548\cite{Kaa_ea00}, 
 unfortunately strongly blended with absorption lines from the 
 Warm Absorber, and perhaps partly produced by it. This continuum 
is the 
``reflected" spectrum, 
 though obviously it is not due to pure reflection! In the hard X-ray range an 
 intense iron line is produced at 6.4 keV, due to Fe~XVII and 
 less ionized species.

\begin{figure}
\begin{center}
\epsfxsize=3.5in 
\caption{Ionization state of iron in the upper layers of an irradiated disk. 
The model is a slab of constant density $10^{14}$cm$^{-3}$, irradiated on 
one side by a power-law $F_{\nu}\propto \nu^{-0.9}$ from 1eV to 100 keV and 
on the other side 
by a BB with $T=10^5$K to mimic the viscous heating. The 
ionization parameter is $\xi=300$, and 
the luminosity of the power-law is equal to the luminosity of the 
BB. 
Computations made with TITAN.}
\label{fig-deg-ion-Fe}
\end{center}
\end{figure}
\begin{figure}
\begin{center}
\epsfxsize=3.5in 
\caption{Emission spectrum for the same model as 
Fig. \ref{fig-deg-ion-Fe}. The dotted line represents the 
underlying transmitted BB
spectrum; the dashed line, the reflected spectrum; and the solid line, the 
total 
spectrum emitted by the disk. The spectral resolution ($R$) 
of the spectrum has been degraded to 30.}
\label{fig-spec-BB-PLx300}
\end{center}
\end{figure}

This model is able to solve several problems raised by the standard 
non-irradiated disk model, like the presence of the soft X-ray excess, 
the iron line and of the ``hump"; but it retains a few unsolved 
problems:
\begin{itemize}
\item Only a small fraction of the UV 
flux  can be 
produced in this way as the X-ray luminosity is 
smaller than $L({\rm BBB})$.
One deduces that the 
optical to UV  emission is only slightly modified with respect to the 
standard disk  (see also Fig. \ref{fig-spec-BB-PLx300}). Since the local 
BB picture holds for the optical to near-UV 
emission, one has
$T_{\rm BB}=[(F_{\rm visc}+F_{\rm X})/\sigma]^{1/4}$, and 
therefore $T_{\rm BB}\propto R^{-3/4}$, except in the very 
central regions, where anyway the BB approximation is not valid and 
only EUV is emitted. 
Consequently to get a steep spectrum, $F_{\nu}\propto 
\nu^{-\alpha}$ with $\alpha\ge 0$ as observed in Seyfert nuclei,
one should seek another geometry for the 
irradiation of disks. 
\item The disk will respond to a 
variation of the central source on a time-scale of
\hbox{$\sim\ 1.2 \lambda_{\mu}^{4/3} \eta^{-1/4}f(R)^{1/4}\dot{m}^{1/4}
M_8^{3/4}$ days}, according to Eq.~(\ref{eq-distancebis}).
Thus, rapid 
X-ray fluctuations will be erased by the light travel-time 
towards the disk, as observed. But the UV--optical light curves should 
always lag behind the X-ray one, while the opposite is sometimes observed, 
for instance in 
the case of NGC 7469. Also the model cannot explain the fact that the UV light 
curve has the same variability amplitude as the X-ray with  the
characteristic UV time-scale.
\item The problem of the Lyman discontinuity remains, although improved with 
respect to non-irradiated disk.
\item Finally, the model requires the 
presence of an ad-hoc X-ray source for 
which no physical emission mechanism is proposed.
\end{itemize}

\subsection{What Can We Learn about the Emitting Medium from the X-ray 
Spectrum?: Where Compton Scattering Makes its Appearence Again} 

We know that the X-ray spectrum is made of a reflected part (the ``cold" iron 
line, the hump), and of a primary spectrum. The reflection takes place 
on a ``cold" 
medium, which most probably is that emitting the UV-bump. The correlated 
variations and the small time-lag of the UV and X-ray fluxes confirm that the two emitting media are 
``seeing" each other. The high energy 
cut-off of the primary 
spectrum at about 100 keV tells us that it is emitted by a hot medium with an electron 
temperature  such that $\Theta_e=kT_e/mc_o^2 \sim 0.1$   (if the 
emission is mainly due to electrons, which is most likely). 

 Three thermal mechanisms can compete in this hot
medium: free-free, cyclotron, and inverse Compton emission.
The time-scale for inverse Compton cooling
is smaller than the time-scale for free-free cooling  when the following 
inequality
holds
\begin{equation}
l ={{L_{\rm UV}} \over R}{{\sigma }_{\rm T} \over m{c}^{3}}\sim 10^4\ 
{L_{\rm UV}\over 
L_{\rm Edd}}{1\over r} > 0.03 \Theta_{e}^{-1/2}\sim 10^{-2}
\mbox{\ \ ,}
\label{eq-Comp-ff}
\end{equation}
where  $l$ is the
``compacity'' of the UV emission region.

The dimension of the UV emission region is 
typically 100$R_{\rm G}$. 
One  deduces that the condition is
easily fulfilled, so inverse Compton
cooling dominates over free-free cooling. It dominates also over cyclotron
cooling unless the magnetic pressure is of the order of the density of 
radiation, which
implies a very high value of the magnetic field.   
	 In conclusion:
{\it the optical-UV and the X-ray emissions are probably
strongly coupled through direct (in the cold medium) and inverse (in the
hot medium) Compton scattering}.

In the hot Thomson 
thick medium
there were multiple scatterings of hard 
photons. In the
 hot Thomson thin medium 
with soft photon input
there is a small number of scatterings. This model was proposed for X-ray 
binaries\cite{SLE76}.

The mean amplification per scattering is $A= 1 + 4\Theta_e - h\nu_0/m_ec^2$.
If $\tau_{\rm T} < 1$, the probability of $m$ scatterings is $\tau_{\rm T}^m$.
After $m$ scatters, $\nu /  \nu_0  = A^m$. 
This creates a  series of humps with decreasing intensities.
One can thus approximate the Comptonized spectrum as a power-law  
$I(\nu) = I(\nu_0) (\nu/\nu_0)^{\ln(\tau_{\rm T})/\ln A}$. 
For a given energy of the soft photons, the spectral index is therefore 
defined  by $\tau_{\rm T}$ and $T_e$. Moreover no photons can be scattered to an 
energy greater than that of the thermal electrons, and thus, one expects the 
power-law to be cut at $h\nu/kT_e \sim 1$. 

Since the cooling of the hot gas is due to inverse Compton scattering, 
its  temperature can be determined from an energy balance 
equation, provided that the heating is known, and thus one can
completely determine the spectral distribution.
This led 
to the introduction
of a 
disk model where the
 release of gravitational energy occurs in a hot optically thin ``corona" above
 the disk, through buoyancy and magnetic field 
reconnection\cite{Lia77}$^,$\cite{HaMa91}$^,$\cite{HaMa93}.
If a large fraction ($f\sim 1$) of the gravitational release
 takes place in the corona,  the disk is 
 almost exclusively heated by the backscattered radiation coming from the corona. It 
 reprocesses  this radiation into a soft BB, and the corona
 upscatters this radiation while being Compton cooled by this 
 process  (cf. Fig. \ref{fig-modeles}).
This radiative coupling between the disk and the corona
and the energy balance of the corona impose a relation
 between $T_e$ and $\tau_{\rm T}$ (the smaller $\tau_{\rm T}$, 
the larger $T_e$), such that 
 the  X-ray spectral index is close to unity. 
 Since an increase in $\tau_{\rm T}$ causes both a decrease 
of $T_e$ and
  a steepening of the spectrum, a relation between the cut-off
 at high energy and the spectral index in the 2-10 keV range is predicted (cf. 
Fig. \ref{fig-haardt-maraschi-97}).

Since the
ratio between the inverse Compton luminosity and the soft luminosity is
of the order of unity,
the model predicts a BBB/X-ray 
luminosity ratio smaller than observed. This is why 
a variant of the disk-corona model, the
``patchy corona" (cf. Fig. \ref{fig-modeles}), was introduced\cite{HMG94}. 
The corona is not homogeneous, but
 is made of a few blobs, 
which could be due to the formation of magnetic 
loops storing energy and releasing it rapidly through reconnection like in 
solar flares.
Below a flare, the disk is heated to a higher temperature.  The rest of the 
disk radiates through viscous processes. An important difference with the 
continuous corona is that the fraction of the disk located below the 
``flare" is much more irradiated and consequently its upper layers are 
strongly modified, in particular due to the large radiative pressure.

\begin{figure}
\begin{center}
\epsfxsize=2.5in 
\caption{Comptonized spectra for different couples of ($\tau_{\rm T},\Theta_e$)
  [(0.63, 0.11), (0.32, 0.21), (0.2, 0.3), (0.1, 0.5)]. The soft photons 
correspond to a BB
 with $kT_{\rm BB}=$100 eV. From Haardt et al.$^{39}$}
\label{fig-haardt-maraschi-97}
\end{center}
\end{figure} 

\subsection{The Two-Phase Disk}

 In the previous model the gravitational release takes place mainly in the 
 corona. 
 This implies strong structural changes  of the disk, both vertically and
 radially, with respect to the 
 standard model.

Disk-corona models have been studied in many  
papers\cite{NaOs93}$^,$\cite{KuMi94}$^,$\cite{ZCC95}$^,$\cite{SvZd94}$^,$\cite{SiKr98}$^,$\cite{Roz_ea99}$^,$\cite{RoCz00}. Since the
 disk has no or only weak internal 
heat generation, it
is  almost isothermal and has a weak radiation pressure. It is therefore 
very dense and 
supported only 
by gas pressure. Consequently {\it it is not subject to the 
viscous/thermal instability like radiation supported disks}. This is 
a great advantage, but on the other hand, since 
these disks are very dense, they are 
gravitationally unstable at all radii for relatively modest luminosities. 
Other differences with 
the lamppost model is that the 
corona pressurizes the disk, and there is conductive transport between the disk
 and the corona.

Generally, the corona is a two temperature plasma, where ions are
heated and transfer their energy to the electrons through Coulomb 
collisions, and the electrons are Compton cooled. The corresponding 
equilibrium temperature is close to virial for the ions (10$^{11}$K), 
and of the order of 10$^9$K for the electrons, as already mentioned. 
Therefore, the gas pressure supported corona is at the limit of being 
geometrically 
thick, and sometimes it is quasi-spherical. Some people think that
a collective 
process able to rapidly transfer energy between ions and electrons can
maintain the two constituents at the same temperature. 

In these models, the fraction of energy 
generated in the corona is
 a free parameter given for instance\cite{KSM00} by a power-law in $R$.
By changing this fraction and the geometry of the
hot region, the
models can fit the observed X-ray spectra.
But one should search for a physical basis for  
the choice. 

Some models take into account mechanical transport only in the 
radial direction, and others add the vertical transport by conduction or 
evaporation between the disk and the corona. 
Models with the local corona strength, calculated on the basis of the
  disk evaporation rate, tend to predict a relatively weak corona above a
  disk, with possible total disk disruption in its innermost 
part\cite{RoCz00}$^,$\cite{MLM00}. 
  These results are supported by 2D numerical time-dependent 
computations\cite{HuCa00}. However, these models
  do not predict strong the soft X-ray excesses present in composite
 spectra\cite{Lao_ea97}, as Comptonization in the corona is too
  weak and the emission from the innermost single phase flow arises in
 the hard X-ray part of the spectrum.

\subsection{The Iron Line: a Proof for Irradiated Cold Relativistic Disks?}

When an X-ray photon is absorbed to eject one of the two K-shell electrons
of an iron ion, this photoelectric process is followed by the 
de-excitation of an
 electron
 from the L-shell, and then, either by the emission of a ``fluorescent" K-photon, 
 or by
a subsequent Auger ionization (escape of an electron from an upper shell without
emission of radiation). The probability of these two processes is of the 
same order and
depends slightly on the ionization state of the ion. The energy of the 
fluorescent photon is almost constant and equal to 6.4 keV for Fe~I to 
XVII, then it increases to 6.9 keV for the H-like Fe~XXVI.  
Actually this transition and the one corresponding to the He-like Fe~XXV 
are not 
``fluorescent"  but simply
 normal resonance L$\alpha$ transitions. Besides, from Li-like (Fe~XXIV) 
 to H-like iron, the Auger effect cannot occur, 
owing to the lack of two
electrons in the L-shell. 
The absorption threshold 
 is 7.1 keV for photoionization of a neutral atom, and it increases 
to 8 keV for Fe~XVII, to reach 9.3 keV for Fe~XXV. 

The ionization state of iron depends on the ionization parameter $\xi$. For 
constant gas density one finds\cite{MFR93}$^,$\cite{MFR96}:
\begin{description}
\item[{\rm for} $\xi \leq 100$:] a ``cold" line is produced  at 6.4 keV by 
ions less ionized than Fe~XVII, and the 
iron K-shell absorption edge is small since the contribution of the other 
elements to the  photoelectric opacity is large.
\item[{\rm for}  $100< \xi \leq 500$:]  iron is in the states 
Fe~XVII to Fe~XXIII. There is a vacancy is the L-shell 
of the ion, so K$\alpha$ 
 photons are resonantly absorbed and trapped until they disappear 
by the Auger effect. Only a few line photons can escape and the Fe~K line is 
very weak. The lower-Z elements are ionized, leading
to a moderate iron absorption edge.
 \item[{\rm for}  $500<\xi \leq 5000$:] the line photons are still subject to resonant
 scattering, but since they are not destroyed by the Auger effect they can escape
 the disk and produce an intense ``hot" iron line
 at 6.8 keV. The absorption edge is strong.
\item[{\rm for} $\xi > 5000$:] iron and the other elements are completely ionized, so 
there is no line and no edge.
\end{description}
As we shall see in the next section, {\it this is no longer valid for a constant 
pressure medium or for a disk in hydrostatic equilibrium}.

The flux of the line and the strength of the ionization edge depend on
 the ionization 
state, but also on the iron abundance, the geometry of the 
reprocessing medium, and, in the case of a plane-parallel slab, on the 
inclination 
of the line of sight and the illuminating source itself. We shall 
not give here more details about these dependences (see 
 Fabian et al.\cite{Fab_ea00}). 
The observed mean intensity, centroid energy, and equivalent width, lead to 
the  conclusion that the line is produced in a relatively ``cold" medium (meaning that 
 iron is less ionized than Fe~XVII), covering about $2\pi$~steradians
of the central 
 source (therefore an infinite slab irradiated by an X-ray source is an 
 acceptable geometry). 

It is widely accepted that the line profile is a proof of the formation of 
the line at the surface of the accretion disk, very close to the BH. 
The extended red wing is then
due to gravitational redshift, while the blue wing is relativisticaly
 boosted due to the
large rotational motion. As an illustration, 
Fig. \ref{fig-profile-iron-K-S} displays on the left the profile of the line emitted 
by a disk around a Schwarzschild BH extending from 6 to 
30$R_{\rm G}$, as a function of the inclination of the disk.
The intensity of the blue wing depends strongly on the 
inclination.
The figure on the right compares the profile of a line formed in an accretion 
disk around a Schwarzschild and a Kerr BH, for an inclination of 
30$^o$. In the case of the Kerr BH, the disk is assumed to extend 
down to 1.25$R_{\rm G}$, so the gravitational redshift is stronger and the red wing is more extended. 

\begin{figure}
\begin{center}
\epsfxsize=2in 
\epsfxsize=2in 
\caption{Profile of the iron line emitted by a disk around a Schwarzschild and a
maximally rotating  Kerr BH. The disk is assumed to extend down to 
the last stable orbit. From Fabian et 
al.$^{31}$, \copyright 2000, Astr. Soc. of the Pacific, 
reproduced with permission of the editors.}
\label{fig-profile-iron-K-S}
\end{center}
\end{figure} 

Fitting the profile of the line in several objects led to a number of
surprising results. First the absence (or the weakness) of
the blue wing corresponds to an inclination close to 30$^o$. Second,
at least in the two best observed objects, 
NGC 3615 and \hbox{MCG-6-30-15}, the 
disk should extend down to 2$R_{\rm G}$ to account for the broad red wings. Fabian et al.\cite{Fab_ea00} argued against  
Reynolds \& Begelman\cite{ReBe97} for a Kerr geometry, 
as the gas falling inside the last stable orbit on a Schwarzschild BH 
 would be highly ionized and the spectrum would display a strong absorption
 edge, inconsistent with the observed data.

\subsection{More Realistic Computations of the Emission Spectrum}

A first attempt to compute the emission of a disk 
topped by a corona in a consistent way was made by Sincell \& 
Krolik\cite{SiKr97}. Their transfer method was already mentioned in the 
discussion of the standard disk. They relaxed the condition of  
constant density, and solved 
 the coupled vertical structure (i.e. hydrostatic equilibrium 
of the irradiated 
layer) and the radiation transfer, assuming that the energy release takes 
place entirely in the corona. The
density in the X-ray heated
skin is very high ($\sim 10^{18}$ 
cm$^{-3}$) and justifies their LTE approach. Integrating the emission over 
the radius, they computed the UV continuum emitted by the  
X-ray heated skin.
Their calculations predict a strong Lyman discontinuity in 
emission, in strong disagreement with the observations.

\begin{figure}
\begin{center}
\epsfxsize=5in 
\caption{Comparison between the temperature structure of an irradiated layer
for constant density and for constant pressure. The curves are labelled with the 
log of the flux incident on the layer. The 
constant density layer and 
the surface of the constant pressure layer have a density 
10$^{12}$ cm$^{-3}$, 
the incident spectrum is a power-law with a spectral index unity, from 0.1 
eV to 100 keV. Computations made with TITAN.}
\label{fig-T-nct-pct}
\end{center}
\end{figure}

Nayakshin and collaborators\cite{NKK00}$^,$\cite{NaKa00} also solved 
consistently the radiation transfer, with a
complete treatment of the ionization and thermal 
equilibrium (not limited 
to H and He) and the hydrostatic equilibrium of an irradiated disk, not 
restricted to the case of a strongly pressurizing corona. They stressed 
that {\it hydrostatic equilibrium is extremely important 
for the structure of the ionized skin}, owing to the
 the thermal 
instability mechanism linked with the $\Xi$ vs. $T$ curve (cf. Fig. 
\ref{fig-Xi-T}): when the temperature reaches the unstable part of the 
curve, it ``jumps" to the other stable part. In contrast with the constant
density case, which 
predicts a smooth decrease of the temperature with increasing depth 
inside the irradiated layer, hydrostatic equilibrium predicts a sharp 
transition between the hot and the cold stable phases. The same phenomenon 
occurs in the constant pressure case, as shown 
in Fig. \ref{fig-T-nct-pct}. One also sees that the hot layer is thicker in 
the constant density case, owing to the rapid increase of density. 
As a consequence, for constant pressure 
the spectrum displays a 
UV-bump restricted to a one temperature BB, while for constant 
density the UV-bump is much more extended in frequency, and it
joins smoothly the
soft X-ray range (cf. Fig. \ref{fig-spe-nct-pct-x3}). 
Fig. \ref{fig-spe-nct-pct-x3-zoom} shows that for
 constant density, no ``cold" iron is present, unlike in the 
constant pressure case, where the cold layers are closer to the surface, 
so the soft photons emitted by these layers are able to leave the medium.

For hydrostatic equilibrium, the emissivity of the line and the spectral 
distribution of the continuum reprocessed in the atmosphere of an
irradiated disk depends  strongly on a dimensionless ``gravitational parameter"
 $A$, proportional to the ratio of the gravity to the radiation 
pressure term\cite{NKK00}$^,$\cite{NaKa00}$^,$\cite{Nay00}. 
Fig. \ref{fig-nayakshin-spectre}
shows that for constant density, when the iron line is intense, it is always 
highly ionized. For hydrostatic 
equilibrium (which isi referred to as the self-consistent 
case), when the iron line is intense,  it 
is a ``cold" line. This might explain the predominance 
of ``cold" iron in AGN spectra. The 
observed correlations are better accounted for by a strong illumination of the 
disk, corresponding to the ``patchy" or ``flaring" corona model, than by 
the lamppost or the continuous corona\cite{NaKa00}. Clearly more of these 
self-consistent computations (unfortunately enormously time consuming) 
are needed
 to get a real physical 
understanding of the complex observed behavior of AGN spectra.

\begin{figure}
\begin{center}
\epsfxsize= 4.5in 
\caption{Comparison between the spectrum reflected by an irradiated layer
for constant density and for constant pressure. Same density and incident 
continuum (the dotted line) as in 
Fig. \ref {fig-T-nct-pct}.}
\label{fig-spe-nct-pct-x3}
\end{center}
\end{figure} 
\begin{figure}
\begin{center}
\epsfxsize=4.5in 
\caption{Zoom of Fig. \ref{fig-spe-nct-pct-x3}. 
Computation made with TITAN and NOAR.}
\label{fig-spe-nct-pct-x3-zoom}
\end{center}
\end{figure} 
\begin{figure}
\begin{center}
\epsfxsize=2.2in 
\caption{Comparison of the constant density case with the hydrostatic 
equilibrium case (``self-consistent case"). The curves are labelled with $\xi$ 
in the first case 
and with the 
gravitational parameter $A$ in the second case. One sees that the iron 
line is ``cold" in the self-consistent case, and ``hot" in the constant density 
case.  From Nayakshin et al.$^{95}$}
\label{fig-nayakshin-spectre}
\end{center}
\end{figure} 

\section{An Alternative to the ``Cold Relativistic Disk"? The ``Cloud Model"}

The iron line and UV continuum are 
both due to reprocessing of the 
X-ray continuum, for instance in the atmosphere of an irradiated disk. Hence
 the regions 
contributing the most to the variable fraction of the UV flux should give rise 
to an iron line, and inversely the region emitting the iron 
line should produce a UV or a EUV excess of continuum emission.

We have seen that the optical-UV variable continuum is emitted at  
large distances from the BH (at $R \sim$ 100 to 1000 
$R_{\rm G}$), with the result that a fraction of the iron 
line is also reprocessed in this region. The iron emission will appear as a
narrow line at 6.4 keV, constant on UV variation time-scales. It is not clear 
whether a narrow core is always observed, but we 
know that the broad-line stays constant 
on time-scales of the order of days while the underlying X-ray 
continuum varies. Also there seems to be a contradiction between the short 
time-lag of the optical and UV light curves and the relative large time-lag between the UV and the X-ray light curves. Generally speaking both the iron
 line  and the 
optical-UV-X observations, in 
particular their variability properties, have
been surprising in the context of the disk model. 

An alternative geometry has been 
proposed\cite{GuRe88}$^,$\cite{CFR92}$^,$\cite{KCR97},
where the Fe~K line
and the 
 UV continuum may arise from dense clouds (optically 
thick for free-free absorption, but not Thomson thick)
with a large covering factor
embedded in a hot quasi spherical medium, rather than from a disk.
A variant of this model,  with Thomson thick clouds,
has also been 
proposed\cite{Col_ea96}.
Fig. \ref{fig-modeles} displays a sketch of the ``cloud model".
In this model
 Compton scattering broadens the iron line. 

A physical basis for the model can be found in a new inhomogeneous 
equilibrium of the disk where the 
bulk of the mass rests in dense clumps magnetically connected in a bath of 
hot  gas\cite{Kro98}.  One advantage of the model is 
that it is thermally and viscously stable. The 
transport of angular momentum is insured by collisions between clumps. The 
requirements for 
dynamical and thermal equilibrium determines the parameters 
of the system, for example the temperature of the hot gas and the optical 
thickness (quite similar to that of the corona in the disk-corona model)
of the
clumps which are Compton 
thick with a covering factor close to unity.

Another possibility exists in the framework of a nuclear wind. 
There are several theoretical arguments for the 
existence of hydrodynamic winds or radiation-driven flows close to the 
BH. On the other hand, the fact that the UV and soft X-ray absorption 
lines are blueshifted is observational evidence for an outflowing medium.
 One can thus imagine that the quasi-spherical medium invoked in this model
is simply the basis of the Warm Absorber, for instance similar to the 
wind envisioned by 
Murray \& Chiang\cite{MuCh95}. One interesting aspect of such a model is
that the iron line would be easily redshifted by the Doppler 
effect from the 
position of the Fe~XXV or Fe~XXVI line (6.7 keV) to 6.4 keV or less, since the 
observer would be seeing  a line emitted only from the illuminated side of 
the clouds. 

 This model is able to account for the UV-X spectrum of 
AGN\cite{Col_ea96}$^,$\cite{CoDu97}$^,$\cite{CzDu98}. 
The observed X-ray continuum is then a mixture of the 
primary 
continuum and of the continuum reflected by the illuminated side of the 
clouds, while the BBB is produced both by intrinsic emission of the 
clouds heated by the X-ray source and by reflection on the illuminated 
side of the clouds. Similar to the disk-corona model,
the X-ray source can be the result of Compton upscatterings of the UV photons 
emitted by the surrounding clouds. The radius of 
the cold medium 
is $r\sim 40 \dot{m} n_{14}^{-1/2}M_7^{-1/2}\xi_3^{-1/2}$, 
where $n_{14}$ is expressed in units of 
10$^{14}$ cm$^{-3}$ and $\xi_3$ in 10$^3$. 
Fig. \ref{fig-spectre-Titan-Noar} shows an example of the spectrum obtained 
for such a model.

\begin{figure}
\begin{center}
\vspace{-2cm}
\epsfxsize=3.8in 
\caption{Spectrum emitted by a spherical shell of clouds surrounding a 
hot medium that upscatters the UV photons emitted by the irradiated clouds. 
The covering factor of the clouds is 0.9, and the ratio of the radius of 
the hot and cold media is 0.4. Spectrum computed with 
TITAN and NOAR. The noise in the 
right-hand side is a numerical artifact of the Monte Carlo method.
The spectral resolution of the left-hand side is 100.}
\label{fig-spectre-Titan-Noar}
\end{center}
\end{figure}

In this closed geometry the UV photons which are 
not upscattered to the X-ray range by clouds in 
the hot medium, undergo multiple  
reflections which amplify the UV-bump. Consequently the clouds ``see" a spectrum 
richer in soft photons than the observer (cf. Fig. \ref{fig-amplification}). 
This leads to a small Compton temperature of the clouds, which is important to 
account for Compton downscattering of the iron line photons (cf. below).

\begin{figure}
\begin{center}
\epsfxsize=3.1in 
\caption{Amplification factor due to multiple reflections in the cloud 
model, assuming that the X-ray source is point-like. Starting from 
$\xi=3000$, the effective primary ionization parameter 
illuminating the clouds reaches 
$\xi=10880$. The covering factor of the clouds is 0.9, and the primary 
spectrum is a power-law with a spectral index equal to unity 
between 0.1 eV and 
100 keV. From Abrassart \& Dumont$^{3}$.}
\label{fig-amplification}
\end{center}
\end{figure} 
\begin{figure}
\begin{center}
\epsfxsize=3.1in
\caption{Fit of the mean iron line profile with the cloud model. The
incident
spectrum on the reprocessing shell is a power-law with unity spectral-index 
between 0.1 eV and
100 keV, the covering factor is 0.9, and the shell contains highly
ionized ($\xi=10000$) and nearly neutral ($\xi=300$) clouds in equal
proportion. From Abrassart \& Dumont$^3$.}
\label{fig-coquille-2xi}
\end{center}
\end{figure}

Such a medium has the power to Compton-broaden the iron line\cite{Cze_ea91}. 
There are actually two possibilities\cite{Mou_ea00}:
\begin{itemize}
\item  the hot medium is surrounded by a cold Compton thick 
medium, so  the iron line and the hump are emitted by the 
Comptonizing medium itself (Comptonization by reflection)\cite{AbDu98};
\item the iron line is intrinsically narrow because it is emitted at $\sim 100
R_{\rm G}$ from the BH, and it is Compton broadened
 when crossing a surrounding ionized 
Compton thick medium (Comptonization by 
transmission)\cite{MiKe98}$^,$\cite{MiSu99}$^,$\cite{AbDu99}.
\end{itemize}

An example of the model fit of the mean AGN spectrum\cite{Nan_ea97}
with Comptonization by reflection is shown in Fig. 
\ref{fig-coquille-2xi}. Two types of clouds are required: ``cold"  clouds, 
which 
emit the ``cold" component of the iron line, and highly ionized 
clouds, which insure Comptonization of the iron line.  The  BBB is produced
by both media\cite{AbDu98}.

An interesting aspect of Comptonization by transmission is that 
it can account not only for the broad red wing of the iron line, but also 
for the absence of a
Lyman discontinuity in the UV spectrum without appealing to 
a fine tuning of the parameters
 (cf. Fig. \ref{fig-transmis-Ledge}).

The transmission model has been 
rejected\cite{Fab_ea95}$^,$\cite{ReWi00}
based 
on the size given by variability. 
The very high ionization state required to avoid an 
intense  iron absorption edge 
implies too large a dimension, in particular 
in the case of \hbox{MCG-6-30-15}.  However, assuming
that the Comptonizing medium is homogeneous (the most conservative case), 
one deduces from Eq.~(\ref{eq-Xi-T}) assuming $\xi=10^6$ and 
 $\tau_{\rm T}\sim 3$ 
a dimension of 0.03 lt-d,  consistent with variability time-scales. 
It  corresponds to a distance of 
about 100$R_{\rm G}$ if the BH in \hbox{MCG-6-30-15} has a mass of a few 
10$^6$M$_{\odot}$, as it is now believed (note that such a small mass implies 
that this object is radiating close to its Eddington luminosity, making it 
peculiar among Seyfert nuclei).  
However a serious argument against the transmission model,
 is that the energy cut should take place at 500 
keV / $\tau^2$, i.e. at about 50 keV, while it seems to be observed above 
100 keV.

\begin{figure}
\begin{center}
\epsfxsize=3.55in 
\caption{Profile of an intrinsic cold narrow 6.4 keV iron line, 
transmitted through an extended shell of Thomson depth 3, with a density 
varying as $1/R^2$ and $\xi=2\times 10^6$. The incident spectrum is a
composite AGN$^{65}$.
 From Abrassart \& 
Dumont$^{4}$.}
\label{fig-transmis-Fe}
\end{center}
\end{figure} 
\begin{figure}
\begin{center}
\epsfxsize=3.55
in 
\caption{Same as in Fig. \ref{fig-transmis-Fe}, but showing the effect on 
an extrinsic Lyman edge in absorption. From Abrassart \& 
Dumont$^4$.}
\label{fig-transmis-Ledge}
\end{center}
\end{figure}

A strong argument against 
the reflection model
is the fact that it 
still has not  been  shown 
able to account for a very extended red wing 
with no blue wing, as observed in \hbox{MCG-6-30-15}. However the 
parameter space is
 far from having been 
completely explored, as these computations are extremely 
time consuming.

The main difference with the disk model 
is that the region where the iron line is 
emitted has not to be located close to the BH. This 
is in agreement with the absence of short term variations of the line 
strength
correlated  with variations of the continuum (though the rapid variations of the 
profile are difficult to account for except with an ad hoc geometry). It 
allows also for the UV and optical emission to arise at similar 
distances, of the order of 100$R_{\rm G}$.  On the other hand, in this model 
the clouds can randomly obscure the line of sight,
thus accounting for short term variations of the primary X-ray 
source\cite{AbCz00}.
Another major difference with the disk model 
is that it also explains naturally 
 a large $L_{\rm UV}/L_{\rm X}$ ratio, owing to the large covering factor of the 
``cold" medium.  Note that this model is in a sense ``non local", as
 the cloud system can be compared to a 
photosphere which is reprocessing the gravitational energy 
that is emitted much closer to the 
BH. 

Finally an intermediate solution has been proposed\cite{Kar_ea00} 
where a system of quasi-spherically randomly keplerian moving clouds 
is located relatively close to the BH. In this scenario the iron line is 
broadened both by relativistic effects and by Comptonization.

\section{And What Happens at the Largest and at the Smallest Accretion 
Rates? The Case for Thick Disks.}

Up to now we have concentrated on accretion rates similar to those observed in 
Seyfert nuclei 
and modest quasars, where
geometrically thin disks are a natural solution of the $\alpha$-prescription. 
It is interesting to consider what happens at both  
higher and lower accretion rates, where thick disks are predicted.

 In thin disks, 
  vertical motions and
radial pressure gradients are  negligible, and it is therefore
possible to neglect
the flux advected towards the center,
\begin{equation}
F_{\rm adv} = {\dot{M}c_{\rm s}^2 \over 4\pi 
R^2}\ f(R) \left({d\ln P\over d\ln R}-{5\over 2} {d\ln T\over d\ln R}\right)
\mbox{\ \ .}
\label{eq-adaf}
\end{equation}
When the disk is geometrically 
thick, the radial velocity is comparable to the rotational velocity, and this 
induces strong radial pressure and temperature gradients. The advected flux 
must then be taken into account,
\begin{equation}
F_{\rm visc}=  F_{\rm cool}+F_{\rm adv} \mbox{\ \ ,} 
\label{equ-Qadv}
\end{equation}
where $F_{\rm cool}$ is the heat flux radiated away. Let us consider the two 
extreme cases:

\begin{enumerate}

\item {\rm for} $\dot{m} \gg {\rm or} \sim  1$:
Owing to the large accretion rate, the flow is dense and optically thick, the
inner regions are thus supported by the strong radiation pressure. 
Consequently, the disk is inflated and becomes geometrically thick.
Heat advected to the
BH exceeds
the heat radiated away, and
 $\eta$ decreases with $\dot{m}$ \cite{ACN80}. For super-Eddington 
accretion rates these disks have relatively small
 luminosities, of the order of $L_{\rm Edd}$.
They radiate 
mainly in the soft X-ray range and it is suggested that they could 
account for Narrow Line Sy~1s, which probably radiate  near their 
Eddington limit, and  display large soft X-ray excess\cite{Min_ea00}.

\item {\rm for} $\dot{m} <  \alpha^2$, {\rm i.e. typically} $\leq 0.01$:
Owing to the small accretion rate,  
the flow has a low density and does not radiate efficiently.
Like the coronae discussed previously, the protons
remain very 
hot and the disk is  supported by gas 
pressure, becoming
geometrically thick and optically thin. 
Several cooling processes can dominate whether or 
not soft photons are 
present to induce inverse Compton cooling\cite{Ich77}$^,$\cite{NaYi95}$^,$\cite{Abr_ea95}$^,$\cite{Nar_ea98}. 
Since these disks are not efficient radiators, a large fraction of the 
gravitational energy will be advected towards the BH and simply ``swallowed".
These disks 
are referred to as ``Advection 
Dominated Accretion Flows" (ADAFs). Both solutions 
(i.e. the standard disk and the ADAF) exist, 
without any understanding on how to switch between them. If the ADAF 
is surrounded by a cold disk the 
emission spectrum will depend on the location of the transition between the 
cold disk and the ADAF (cf. Fig. 
\ref{fig-ADAF}). In this case 
the hot medium constituting the ADAF can be cooled by inverse Compton 
scattering of the soft photons produced by the cold disk. If a 
nuclear radio source is present, the synchrotron photons can play the same 
role in cooling the hot gas.

\end{enumerate}

\begin{figure}
\begin{center}
\epsfxsize=15pc 
\caption{Spectrum of an ADAF with a cold disk. The 4 curves correspond to 
different positions of the transition between the ADAF and the cold 
disk: infinity (no cold disk, pure flat ADAF 
spectrum), 1000 $R_{\rm G}$, 100$R_{\rm G}$, 10$R_{\rm G}$. The importance of the blue bump and 
its peak frequency increase when the transition radius decreases. Courtesy 
J.-P. Lasota.}
\label{fig-ADAF}
\end{center}
\end{figure}

Much recent attention has 
been devoted these last years 
to ADAFs. They are generally thought to be present in low 
luminosity objects, such as LINERs and the nuclei of elliptical 
galaxies (M~87\cite{Rey_ea96}), the nucleus of 
NGC~4258\cite{Las_ea96}, and the Galactic 
Center\cite{MDZ96}. All show a non-resolved X-ray source 
but no optical/UV continuum, as one expects 
from an ADAF.  The theoretical possibility of 
an ADAF is a matter of current debate, and observationally it is controversed 
(in particular for Sgr A*\cite{Ago00}).

\section{A ``No Man's Land": the Disk between 10$^3$ and 10$^6\ R_{\rm G}$}

We have seen that the disks become 
gravitationally unstable beyond 
a few 10$^3 R_{\rm G}$. 
Moreover, the 
viscous time-scale given by the $\alpha$-prescription for transporting the gas towards
 the BH at such 
distances is
larger than the lifetime of the active nucleus.  We know that further from 
the  center, at about 10$^6\ R_{\rm G}$, the disk is
globally gravitationally unstable and the supply of gas can be achieved by 
gravitational torques (cf. F.Combes in these lecture notes). However we know 
nothing about the structure of the disk in the intermediate region and how 
the outward 
transport of angular momentum is ensured.

Several solutions have been proposed. A magnetic field anchored in the disk 
can transport angular momentum non-locally, and possibly prevent 
gravitational instability. Another solution consists of
a disk made from  marginally unstable and
randomly moving clouds with large bulk velocities, where the transfer of angular
 momentum is provided by cloud 
collisions\cite{BFS89}. A more elaborate model 
is proposed for NGC 1068\cite{Kum99} in which a 
disk of gas clumps undergoes mutual gravitational interactions. 
These disks would be geometrically thick, and 
are actually similar to the tori invoked in the Unified Scheme.
 
The observations of purely keplerian velocities in masers 
lying at about 10$^5\ R_{\rm G}$ from the BH in NGC~4258\cite{NeMa95}
argue strongly for a geometrically thin warped disk.
Collin \& Zahn\cite{CoZa99} have therefore adopted the view of a 
marginally stable disk where unstable 
clumps
collapse until protostars are formed. (However, such a picture is 
only well suited for low Eddington ratios. For large Eddington ratios 
the number
 of newly
 formed stars is high, and the 
disk will become highly
turbulent and inhomogeneous.)
Stars embedded in an accretion disk
should accrete at a high 
rate and rapidly acquire masses of a few 
tens of M$_{\odot}$, leading to a ``starburst" of massive stars (not 
comparable in size with real starbursts)\cite{ALW93}. 
The mass transport can thus be driven by supernovae or by stellar 
outflows, which at the same time 
can
contribute to the heavy element enrichment observed in the BLRs at 
high-$z$ and in BALs. A larger scale starburst 
could also be induced in the shocked gas of  
supernova remnants.   Several pieces of evidence
have been discussed in these lectures for the existence of  
starbursts in the very central region of Seyfert galaxies. 

Finally Ostriker\cite{Ost83} made a very
 interesting suggestion 
that has never been seriously considered.  He stated that if 
the disk is embedded in a 
massive stellar system, interactions between the stars and the disk can
remove angular momentum from the disk at the rate expected for an AGN.

\section{Conclusion}

It is not easy to conclude such a chapter because the results are really 
difficult to summarize. 
We have seen that the tremendous increase of data with the advent of 
X-ray missions such as  ASCA, RXTE, and now Chandra and XMM, has 
considerably complicated the problem, and 
now precludes any simple interpretation.  We have seen that the most 
detailed models, used to make 
observational predictions, are still based on 
crude approximations, such as geometrically thin, stationary and axisymmetric 
disks. For instance the models are generally assumed to be steady-state, 
despite the evidence that there exist 
strong
 variations on small time-scales. The viscosity is currently 
parametrized by
 the $\alpha$-prescription, but other prescriptions are not excluded. 
There are heated debates about the value of $\alpha$, whether it is 
constant or not, whether the stress tensor is  proportional to the total 
pressure or to 
the gas 
pressure. We don't know the influence of the magnetic field.  
We have no idea of the geometry of 
the inner regions. We know that there is a quasi-spherical hot accretion flow, 
but at the same time we believe that
the cold thin disk extends down to a few gravitational radii from the BH,
 and in this case 
we know that it should be unstable. The status of the hot flow is not 
clear yet. 
In short, one could say that the more 
 sophisticated 
 the models, the larger the uncertainties! To continue to make progress, 
we must obtain very 
good new observational data, and at the 
same time develop the basic physics and tools for the modeling. The field has 
still a promising future in the third millenium!

\section*{Acknowledgments}

I am grateful for enlighting discussions with B. Czerny, A-M. Dumont, J-M. 
Hur\'e,  M. Mouchet, and J-P. Zahn and to I. Aretxaga for a careful 
reading of the 
manuscript, leading
to many clarifications and improvements.

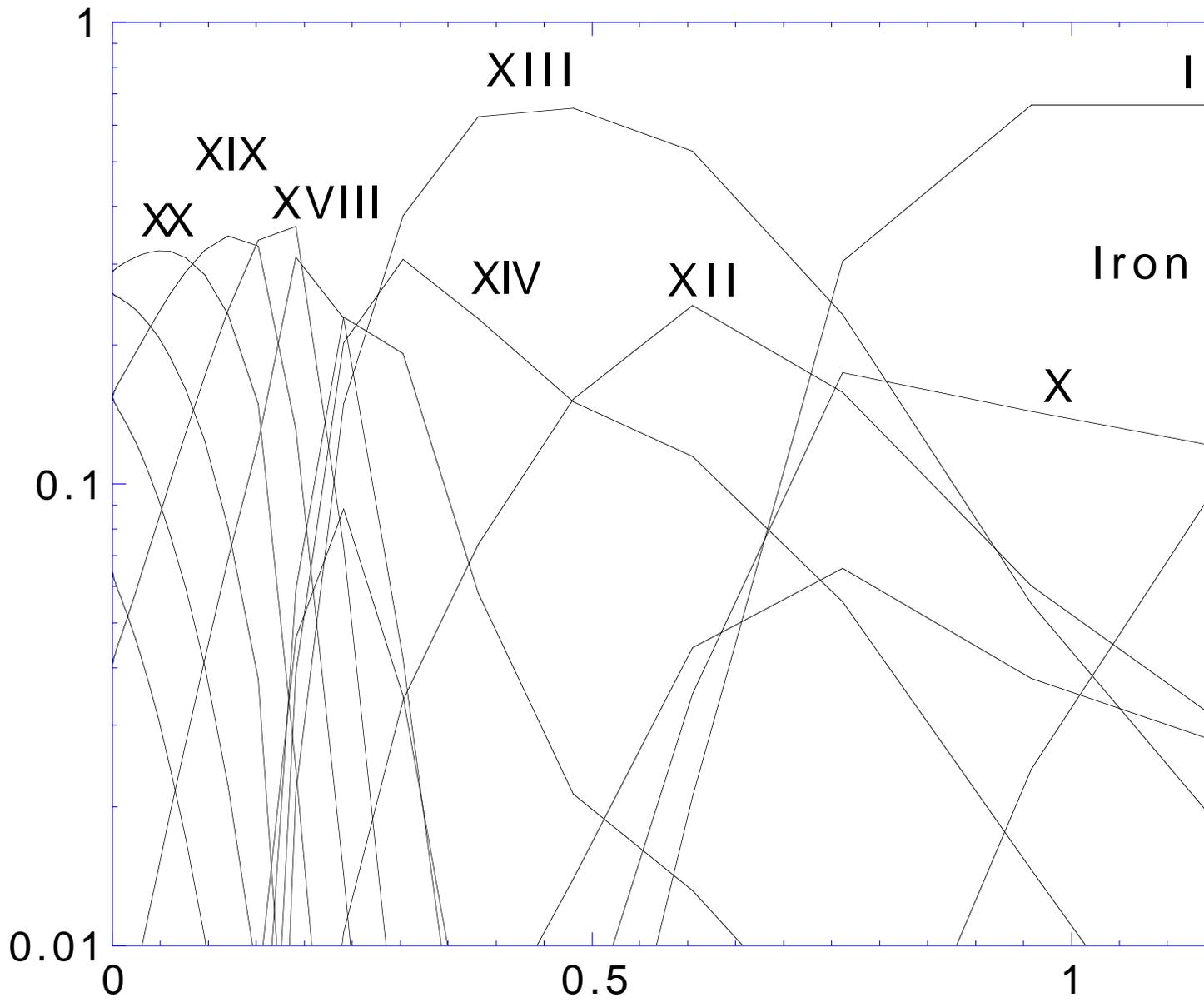
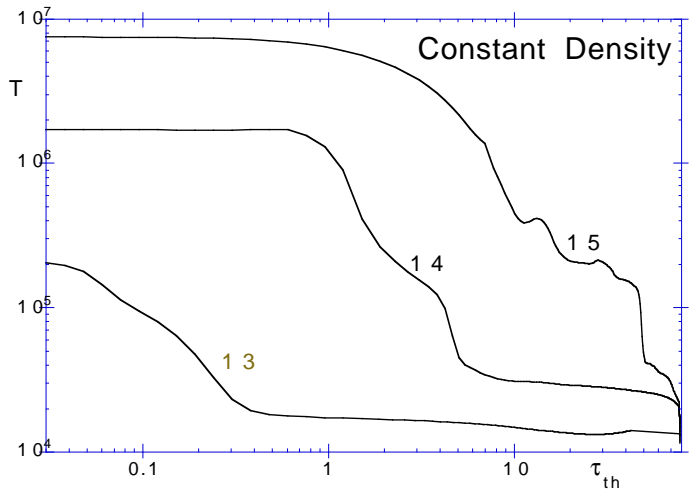
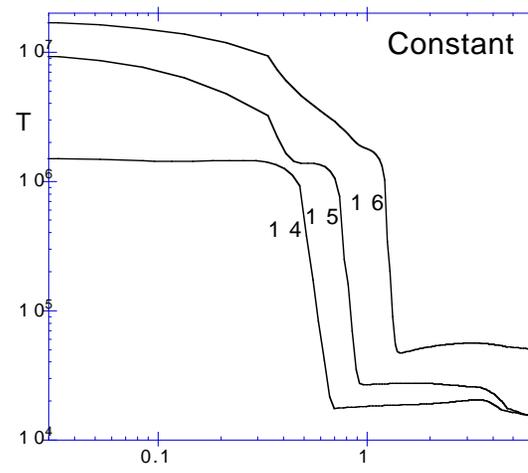
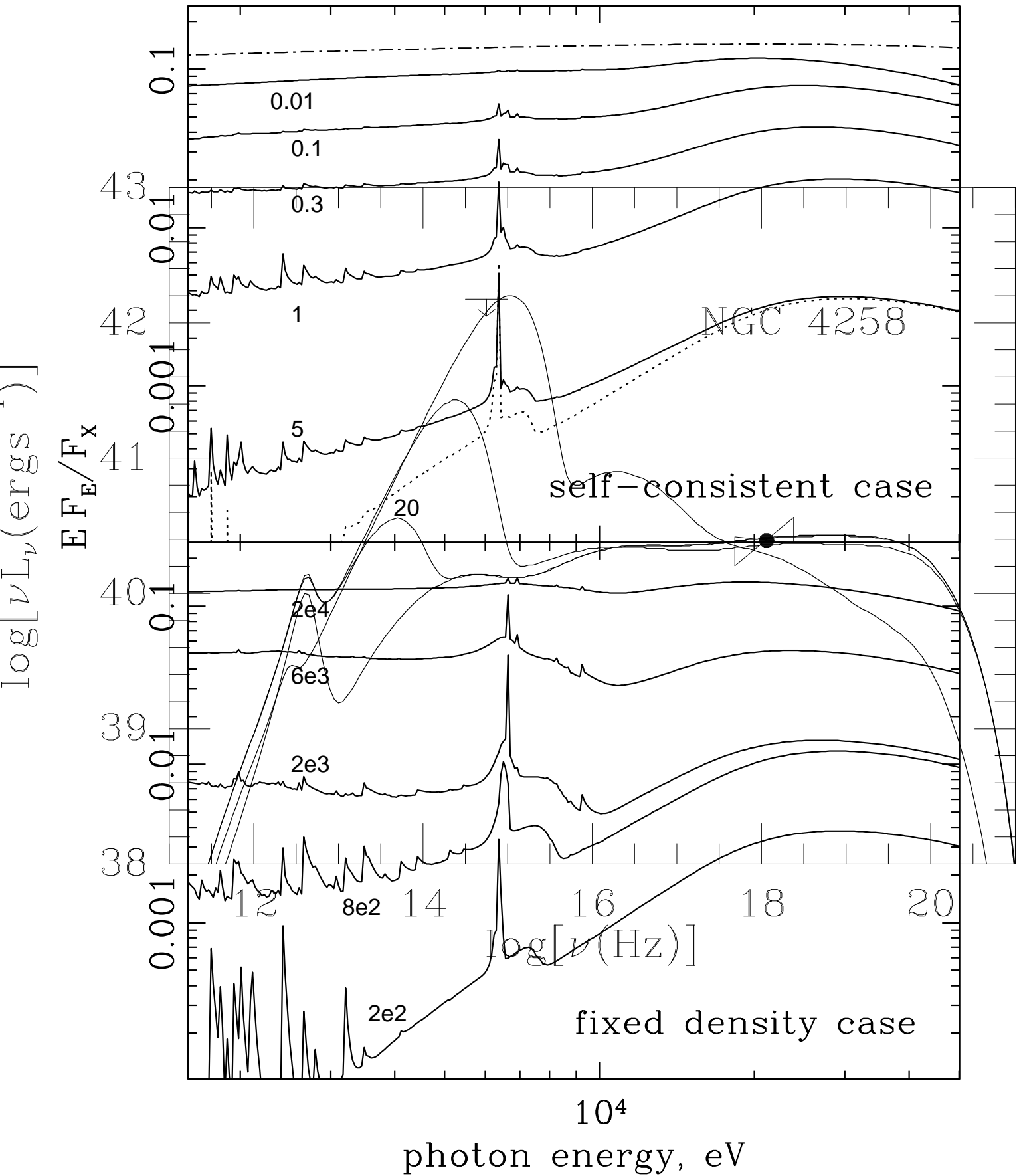
\end{document}